\documentclass[aps,prl,reprint]{revtex4-1}
\usepackage{amsmath}
\usepackage{amssymb}
\usepackage{graphicx}
\usepackage{mathrsfs}
\usepackage{subfigure}

\newcommand{ \braces} [1] { \left( #1 \right)}
\newcommand{ \ebraces} [1] { \left[ #1 \right]}
\newcommand{\von} [1] {\! \braces{#1}}

\begin{document}

\title{Engineering separatrix volume as a control technique for dynamical transitions}

\author{Timo Eichmann}
\author{Eike P. Thesing}
\author{James R. Anglin}

\affiliation{\mbox{State Research Center OPTIMAS and Fachbereich Physik,} \mbox{Technische Univerit\"at Kaiserslautern,} \mbox{D-67663 Kaiserslautern, Germany}}

\date{\today}

\begin{abstract}
Dynamical transitions, such as a change from bound to unbound motion, often occur as post-adiabatic crossings of a time-dependent separatrix. Whether or not any given orbit will include such a crossing transition typically depends sensitively on initial conditions, but a simple estimate for the fraction of orbits which will cross the separatrix, based on Liouville's theorem, has appeared several times in the literature. Post-adiabatic dynamical transitions have more recently been reconsidered as a control problem rather than an initial value problem: what forms of time-dependent Hamiltonian can most efficiently induce desired transitions, or prevent unwanted ones? We therefore apply the Liouvillian estimate for the transition fraction to show how engineering separatrix volumes in phase space can be a control technique for dynamical transitions.
\end{abstract}

\maketitle

\section{I. Introduction}
Hamiltonian systems may exhibit multiple dynamical phases, with qualitatively different kinds of time evolution occurring in different regions of phase space. Basic examples are the different phases of bound and unbound motion for a particle in a finite potential well, as well as the spinning and oscillating phases of a physical pendulum. In such examples, a time-independent Hamiltonian will never let the system cross the separatrix between the two phases, but dynamical transitions can occur when Hamiltonians are time-dependent. Since the concept of a dynamical transition implies a time scale hierarchy that allows us to distinguish the two distinct phases of motion from the transition between them, dynamical transitions have long been studied as post-adiabatic effects under slowly time-dependent Hamiltonians.
\cite{Kruskal,Neishtadt1,Timofeev,Henrard,Tennyson,Hannay,Cary,Eiskens,Neishtadt2,Chernikov,bucket,proof,Bazzani,Jarzynski}

\begin{figure}[htb]
	\centering
	\includegraphics[width=.475\textwidth]{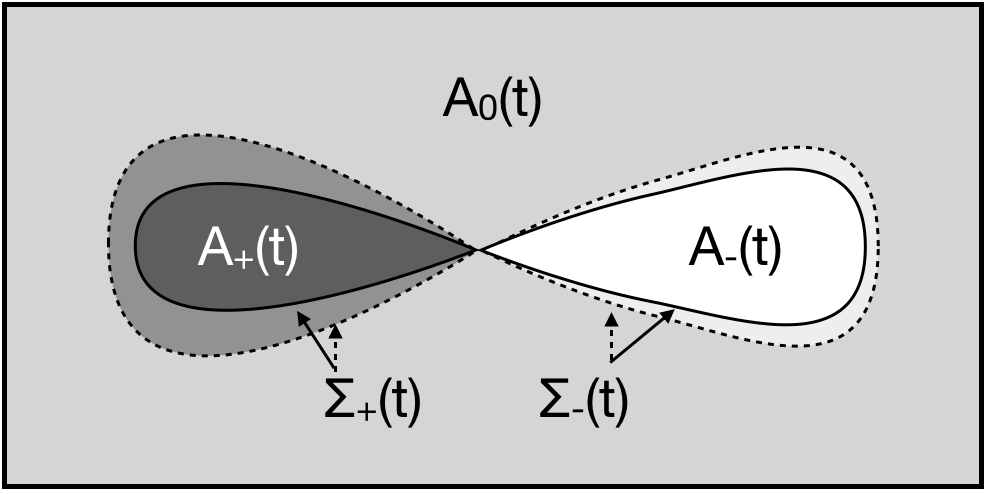}
	\caption{\label{fig:sepsketch} Sketch of phase space regions exhibiting different dynamical phases, divided by two separatrices $\Sigma_{\pm}$ into three regions shaded dark, gray, and white, with areas $A_{+}$, $A_{0}$, $A_{-}$ respectively. The dashed curves and lighter shading indicate that the separatrices move and grow during a short interval $\delta t$ within which the Hamiltonian is time-dependent. According to the theory presented in the text, the rates $\dot{A}_{\pm}$ and $\dot{A}_{0}\equiv -(\dot{A}_{+}+\dot{A}_{-})$ at which the three phase space areas change determine an adiabatic estimate for the fractions of trajectories that will be drawn from $A_{0}$ into either $A_{+}$ (hypothetically, a desired goal) or $A_{-}$ (the unwanted alternative) during the interval $\delta t$.}
\end{figure}

A paradigmatic example is shown in Fig.~\ref{fig:sepsketch}. Three dynamical regions in phase space are separated by two instantaneous separatrices $\Sigma_{\pm}(t)$ which depend slowly on time $t$ because the Hamiltonian $H$ does. According to the adiabatic theorem, orbits sufficiently far from any separatrix will remain within the same dynamical region, but the adiabatic approximation breaks down near a separatrix. As the separatrices slowly move and deform, therefore, orbits can spill through them from one dynamical phase into another, because the orbits will be non-adiabatic when a separatrix is near.

Even numerical results for such problems can be difficult to obtain with high precision \cite{Jarzynski}. Systematic `neo-adiabatic' treatments have been developed to determine the changes in adiabatic invariants over the non-adiabatic interval of a separatrix crossing by combining three different kinds of approximation whose zones of applicability overlap in phase space \cite{Timofeev,Tennyson,Hannay,Eiskens}.  Separatrix crossing has more recently been re-examined, however, from the perspective of control theory \cite{bucket,Bazzani}, where instead of solving the initial value problem for a given time-dependent Hamiltonian, one asks rather what kind of time-dependent Hamiltonian may generate the time evolution which most efficiently achieves a particular goal, for some given set of initial conditions whose preparation is feasible.

Fully detailed post-adiabatic calculations are likely to be too long and complex to be convenient guides for Hamiltonian engineering in systems with many tunable parameters. From the point of view of control, though, the most important issue in separatrix crossing is simply the fraction of preparable initial states for which the desired transition occurs. In this paper we therefore point out that this important question of transition probability can be answered by using only one small and simple part of adiabatic theory, namely a formula based on Liouville's theorem \cite{Goldstein} that was originally presented by Kruskal, Neishtadt and Henrard and has since been extended by others.

We will begin in Section II below by presenting this basic idea for estimating transition probability from Liouville's theorem, but then also critiquing it with some apparent numerical counterexamples. In Section III we will explain how the simple formula can be improved and extended, and conclude that simple Liouvillian estimates for probabilities of dynamical transitions really can be robustly accurate. In Section IV we will apply this picture to some transition crossing problems that have been posed as control tasks; in particular we will provide an analytical theory that accurately explains some recent numerical data presented in \cite{Bazzani}. We will conclude in Section V with a discussion comparing the ``Liouville control'' principle of engineering separatrix volume growth to the thermodynamic requirement of entropy increase, as conditions for spontaneous change.

\section{II. Transition probabilities from Liouville's theorem}
\subsection{A. The Kruskal-Neishtadt-Henrard formula}
The simplest example of quickly deducing probabilities for crossing separatrices can be illustrated with Fig.~\ref{fig:sepsketch}. Suppose that both separatrices $\Sigma_{\pm}$ steadily grow during the interval from time $t$ to $t+\delta t$, so that each $\Sigma_{\pm}(t+\delta t)$ entirely encloses the $\Sigma_{\pm}(t)$, as suggested in Fig.~\ref{fig:sepsketch}. If $A_{\pm}(t)$ and $A_{0}(t)$ denote the respective phase space areas of the three regions into which the separatrices divide all of phase space, then $A_{+}$ and $A_{-}$ are growing while $A_{0}$ is correspondingly shrinking. Phase space orbits must therefore be crossing post-adiabatically from $A_{0}$ into $A_{+}$ and $A_{-}$ during the interval $\delta t$. Kruskal, Neishstadt and Henrard have all appealed to Liouville's theorem to deduce that the fraction of these separatrix-crossing orbits which enter $A_{+}$ or $A_{-}$ respectively must be
\begin{equation}\label{Pn00}
\mathcal P_{\pm}=\frac{\delta{A}_{\pm}}{\delta{A}_{+}+\delta{A}_{-}}\equiv-\frac{\delta{A}_{\pm}}{\delta{A}_{0}},
\end{equation}
where $\delta A_{\pm}$ denote the changes in area of the respective regions over $\delta t$. 

In the limit $\delta t\to0$ so that $\delta A_{\pm}\to \dot{A}_{\pm}\delta t$ we obtain an essentially equivalent expression which can be interpreted as the probability that an orbit will enter $A_{+}$ or $A_{-}$, given that it moves into one or the other of them from $A_{0}$ at time $t$:
\begin{equation}\label{Pn0}
\mathcal{P}_{\pm}(t)=\frac{\dot{A}_{\pm}}{\dot{A}_{+}+\dot{A}_{-}}\equiv-\frac{\dot{A}_{\pm}}{\dot{A}_{0}}.
\end{equation}
If the explicit time dependence of the Hamiltonian is slow, then the three areas and hence also $\mathcal{P}_{\pm}$ vary only slowly with $t$, and so (\ref{Pn0}) can be applied to any set of trajectories which all choose between $A_{+}$ and $A_{-}$ at around the same time $t$, without having to determine exactly when or where any particular orbit will meet a separatrix, as long as one can invoke a certain weak kind of ergodicity to assume that the set of trajectories is typical of all those that move from $A_{0}$ into $A_{\pm}$ in the time around $t$.

Equation (\ref{Pn0}) may at first seem a strange proposition for deterministic mechanics. It speaks of probabilities. Yet once formulated, it hardly even seems to need proof. Liouville's theorem tells us that time evolution in phase space is an incompressible flow, even when Hamiltonians are time-dependent. The total increase in separatrix area $\delta A_{+}+\delta A_{-} = -\delta{A}_{0}$ therefore represents a certain conserved volume of possible system orbits which have entered one or the other separatrix during the time-dependent interval. The two area increases $\delta A_{+}$ and $\delta A_{-}$ are conserved measures of the number of orbits which have entered each individual separatrix. If we know that a given orbit is within the entering set of measure $-\delta A_{0}$, therefore, and if that is all that we know, then (\ref{Pn0}) is the obvious guess for how likely it is the orbit ends up inside $\Sigma_{+}$ or $\Sigma_{-}$ in particular. 

The formula can also be extended straightforwardly to cases where one of the `destination' areas $A_{\pm}$ also shrinks over time, as $A_{0}$ does, instead of growing. If $A_{+}$ is shrinking then all orbits that remain inside $\Sigma_{+}$ adiabatically must be ones that were already there, and so no additional room is available for any new orbits to enter the shrinking $A_{+}$ through the incompressible flow of Hamiltonian time evolution. Hence if $\delta A_{+}<0$ we conclude that $\mathcal P_{+}=0$. If on the other hand $A_{-}$ and $A_{0}$ are both shrinking while $A_{+}$ grows, then the same combination of adiabatic and Liouvillian reasoning implies that orbits which migrate to a new region must all migrate to $A_{+}$, and so $\mathcal P_{+}=1$. In other words, if formula (\ref{Pn0}) yields a result greater than one or less than zero, it is to be interpreted as one or zero, respectively. 

The formula (\ref{Pn0}) was published without proof by Dobbrot and Greene in 1971 \cite{Kruskal}, where it was attributed to Kruskal in a private communication, referred to as ``Kruskal's theorem," and used to examine motion of charged particles in a class of magnetic confinement devices (``stellarators'') intended for fusion power generation. A derivation of this formula was given by Neishstadt in 1974 \cite{Neishtadt1}, motivated by questions about orbital resonances among the moons of Saturn. An independent derivation of (\ref{Pn0}) was added in 1982 by Henrard \cite{Henrard}. We will therefore refer to (\ref{Pn0}) as the Kruskal-Neishtadt-Henrard formula (KNH).

This somewhat abstruse history of the KNH formula may well make a reader think again about just how obviously valid the formula is. On second thought, in fact, the KNH formula may become downright dubious. It attempts to draw conclusions about how frequently an essentially \emph{non-adiabatic} phenomenon will occur, based on geometrical quantities that are only defined \emph{adiabatically}. Although Liouville's theorem is exact, time-dependent separatrices are really only instantaneous separatrices within the adiabatic approximation, and so the KNH formula can only be as good as that approximation; and yet it assigns probabilities to different cases in which the adiabatic approximation breaks down.

The question of whether the KNH formula is valid is not just an academic paradox. The formula offers a way to determine an important feature of non-adiabatic evolution---namely the probability of a separatrix-crossing transition---merely by determining the instantaneous separatrices of the Hamiltonian $H(t)$, without having to solve for the actual system time evolution. Having an estimate of transition probability without solving non-adiabatic evolution may be merely a convenience, if one is trying to solve the initial value problem for a given Hamiltonian, as for instance to predict the motion of satellites. If however one instead faces the control task of getting the system into $A_{+}$ from $A_{0}$, and if one's means for achieving this task are various ways of modifying $H(t)$, then a way of estimating the chance of success from the instantaneous Hamiltonians alone may be more than just convenient. It may enable one to replace trial and error with deliberate design. 

How generally valid is (\ref{Pn0}), even when time dependence is slow? Careful examination shows that the formula stated in (\ref{Pn0}) can in general fail badly. Some simple examples will illustrate the problem, but then also suggest a solution.

\subsection{B. Application of the KNH formula}

The concrete application of the intuitive KNH formula, as well as its limitations, can both be seen by re-examining a specific example that has been considered by many authors including both Neishtadt and Henrard, namely the Hamiltonian
\begin{equation}\label{E:pertPendH}
	H_1=\frac{\big(P-\alpha(t)\big)^2}{2}-\beta^{2}(t)\cos(\phi),
\end{equation}
with $\phi$ and $P$ canonically conjugate coordinates, and $\alpha$ and $\beta$ slowly changing parameters. This model has been studied in a wide range of physical contexts, but as a simple concrete realization one could consider $\phi$ and $P$ to be the one-dimensional position and momentum, respectively, of a charged particle in an electric field which is a superposition of spatially sinusoidal component and a spatially constant component, each component being of time-dependent strength. This electric field is represented in a gauge such that the sinusoidal component is due to the electrostatic potential while the constant component is due to the vector potential $\propto \alpha(t)$.

This Hamiltonian (\ref{E:pertPendH}) has two time-dependent instantaneous separatrices $(\phi,P)\to(\phi,P_{\pm}(\phi,t))$ dividing phase space into three parts:
\begin{equation}\label{seporg}
P_{\pm}(\phi,t)=\alpha(t)\pm 2\beta(t)\cos\frac{\phi}{2}\;.
\end{equation}
The geometry is different from Fig.~\ref{fig:sepsketch} because of the periodicity of $\phi$; see Fig.~\ref{H1sep} (a simple sketch for comparison with Figs.~\ref{fig:sepsketch} above and \ref{H2sep} below) as well as the two upper panels of Fig.~\ref{fig:sep} (more detailed plots showing several energy contours, which the system follows adiabatically in its time evolution). 
\begin{figure}[htb]
	\centering
	\includegraphics*[width=0.45\textwidth]{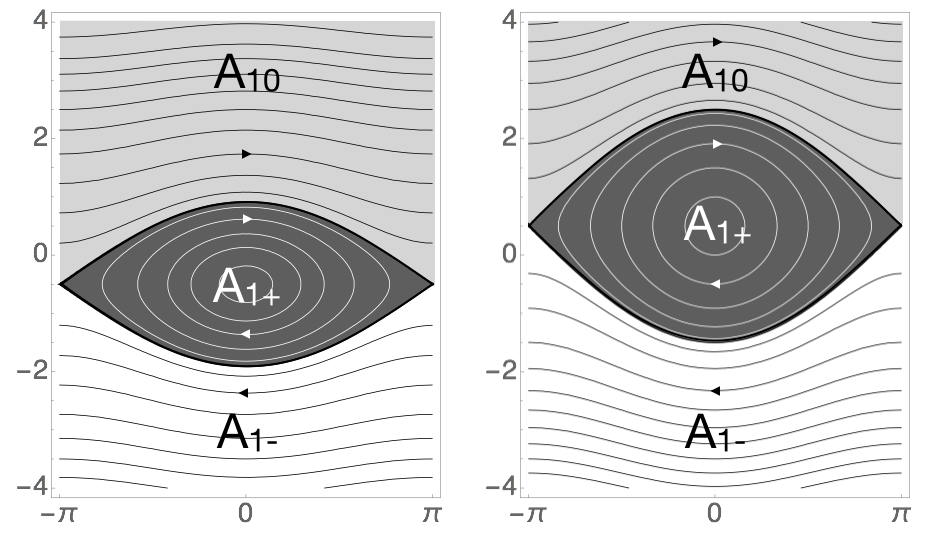}
	\caption{\label{H1sep}Contours of constant $H_{1}$ in the phase space of $\phi$ (horizontal axis) and $P$ (vertical axis) for example cases of $H_{1}$ with $(\alpha,\beta)$ given by (-1/2,1/2) (left panel) and (1/2,1) (right panel). The contours drawn in thicker black line are the separatrices, which divide the full phase space into three regions that correspond conceptually to the three regions of Fig.~\ref{fig:sepsketch}. In the adiabatic limit the system's time evolution is to flow along these energy contours, in the directions indicated by the small arrowheads in each region; post-adiabatic corrections make actual orbits drift to different energy contours and even move between regions. The $\phi$ coordinate is periodic, such that motion wraps from the right edge of each frame to the left edge. Here the left and right panels could represent the same time-dependent Hamiltonian at different times, showing how all three areas $A_{1\pm,0}$ can change in time, although their sum remains constant. }
\end{figure}

We will take the finite region between the two separatrices to be our target region $+$; the area $A_{1+}$ is then easily computed as
\begin{equation}\label{separea}
A_{1+}(t)=\int_{-\pi}^{\pi}\!d\phi\,\left(P_{+}(\phi,t)-P_{-}(\phi,t)\right)= 16\beta(t)\;.
\end{equation}

We can define finite areas $A_{10}$ and $A_{1-}$ above and below the two separatrices by setting upper and lower boundaries in $P$ that are far enough away from the separatrices, throughout the entire time evolution of interest, that orbits near them have essentially constant $P$ and so no orbits will ever exit above $A_{10}$ or below $A_{1-}$. We can then also compute
\begin{align}
A_{10}(t) &= \bar{A}_{10} - \int_{-\pi}^{\pi}\!\!\!d\phi\,P_{+}=\bar{A}_{10} - 2\pi\alpha(t) - 8\beta(t)\nonumber\\
A_{1-}(t) &= \bar{A}_{1-} + \int_{-\pi}^{\pi}\!\!\!d\phi\,P_{-}=\bar{A}_{1-} + 2\pi\alpha(t) - 8\beta(t)
\end{align}
where $\bar{A}_{10}$ and $\bar{A}_{1-}$ are arbitrary large constants determined by exactly where we place our constant $P$ boundaries to the outer regions.

With Neishtadt and Henrard we take examples with $\dot{\alpha}>0$, and we consider orbits which begin in the $A_{10}$ region above the separatrices. These orbits will eventually encounter the separatrix $P_{+}$ at some time $t$ (or any time near this $t$, since the separatrices move only slowly), and then either be captured into $A_{1+}$, or else emerge into $A_{1-}$. According to \cite{Neishtadt1} and \cite{Henrard}, and from applying our (\ref{Pn0}), the adiabatic approximation for the fraction of such orbits that will end up inside $A_{1+}$ will therefore be
\begin{equation}\label{Kruskal1}
\mathcal P_{1+}(t) = \left\{\begin{array}{lcl}0&,&\dot{\beta}\leq 0\\ \frac{8\dot{\beta}}{4\dot{\beta}+\pi\dot{\alpha}}&,&\dot{\beta}>0,\;\dot{\alpha}\geq 4\dot{\beta}/\pi\\
1&,&0<\dot{\alpha}<4\dot{\beta}/\pi\end{array}\right.\;.
\end{equation}

\subsection{C. Limitations of the KNH formula}

The fact that Eqn.~(\ref{Kruskal1}) as stated is not generally accurate, however, can be seen by numerically solving cases with time-independent $\beta=1$. In all such cases $\dot{A}_{1+}=0$, because the only time dependence of the separatrices is a rigid translation in $P$ by $\alpha(t)$, as shown in the top panels of Fig.~\ref{fig:sep}, and so $\mathcal P_{1+}=0$ according to (\ref{Kruskal1}). 
\begin{figure}[htb]\label{fig:sep}
	\centering
	\includegraphics*[trim=50 10 30 10,width=0.5\textwidth]{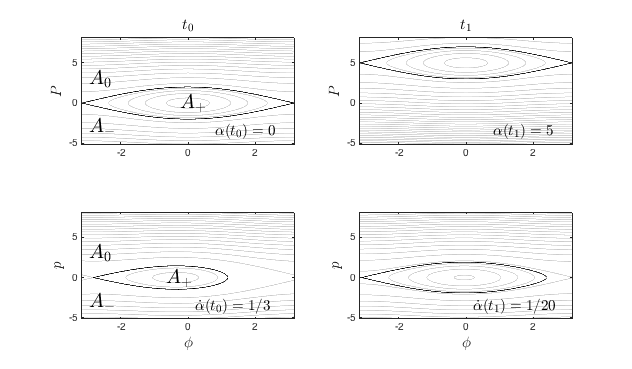}
	\caption{\label{fig:sep} Separatrices and other energy contours for $H_{1}$ (top panels) and for the canonically equivalent $H_{2}$ (bottom panels) at two different times $t_{0}$ and $t_{1}$ (left and right panels). The parameter $\beta=1$ is here constant at all times, and $\alpha(t)$ changes such that $\alpha(t_{0})=0$ and $\alpha(t_{1})=5$, while the rates of change are $\dot{\alpha}(t_{0})=1/3$ and $\dot{\alpha}(t_{1})=1/20$. Although the two Hamiltonians are canonically equivalent, under $H_{1}$ the region $A_{1+}$ that is enclosed by the separatrix moves but does not grow, while under $H_{2}$ the corresponding $A_{2+}$ grows without moving.}
\end{figure}

\begin{figure}[!h]
\centering
\includegraphics*[trim=130 30 150 40 ,width=.475\textwidth]{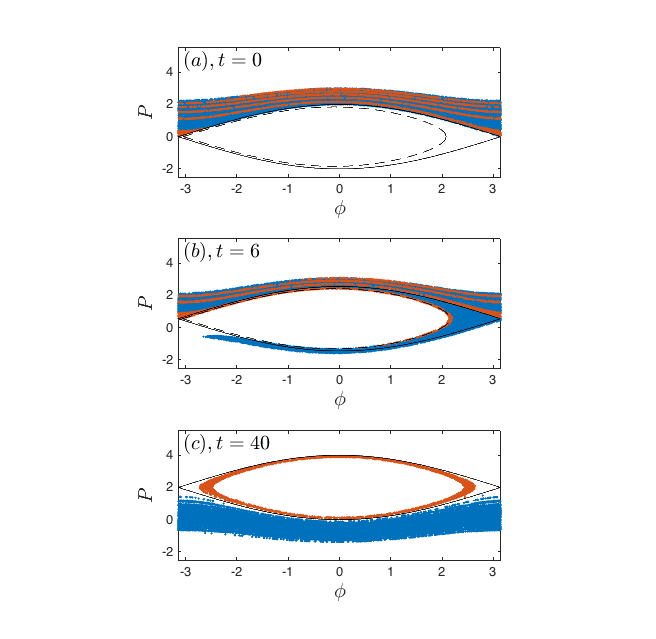}
\caption{\label{fig:decel} \textit{Capture fraction}. 20000 initial conditions (a) evolved to two later times (b) and (c), under the canonical equations generated equivalently by $H_{1}$ of (\ref{E:pertPendH}) and $H_{2}$ of (\ref{Htrans}), with $\beta(t)=1$ and $\alpha(t)=(t/10)(1-t/80)$. Solid black curves are the instantaneous separatrices according to $H_{1}$; dashed curves are the $H_{2}$ separatrices; the two sets of separatrices coincide at $t=40$ (c). The approximately 15\% of points that will be inside $A_{1+}$ in (c) are shown at all times in red; other points are in blue. Applying the KNH formula using the areas enclosed by the solid $H_{1}$-separatrices incorrectly predicts zero capture fraction, while an improved formula based on $H_{2}$ provides an accurate estimate.}
\end{figure}

As Fig.~\ref{fig:decel} shows, however, a particular case in which $\ddot{\alpha}(t)<0$ turns out to have a capture fraction into region $A_{1+}$ of about 15\%---high enough that if this evolution represented a chemical reaction with a valuable product \cite{Cary} it might be considered an acceptable yield. The KNH formula using the separatrix areas for the Hamiltonian $H_{1}$ has completely failed to predict this significant outcome.

For another perspective on how (\ref{Kruskal1}) fails we can consider an alternative control task: instead of capturing the system into $A_{1+}$, we now wish to keep the system inside $A_{1+}$ in order to transport the system to a substantially higher value of $P$ \cite{bucket,Bazzani}. The logic behind (\ref{Pn0}) and (\ref{Kruskal1}) implies for this case that orbits which are initially in $A_{1+}$ will all remain there, as long as $A_{1+}(t)$ never shrinks. As reported in \cite{Bazzani}, however (and discussed here in Section IV below), a significant fraction of such initial orbits fail to be transported significantly because they escape from the separatrix.

In both these cases the KNH formula fails qualitatively. What has gone wrong with it? Was it never really anything more than a hand-waving argument which appeared universal because it invoked Liouville's theorem but which is unfortunately ruined by invalid application of adiabatic approximations?

The KNH formula's basis in Liouville's theorem about incompressible phase space flow is truly strong, since Liouville's theorem is exact even for time-dependent Hamiltonians \cite{Goldstein}. The problem does indeed lie in the consideration of areas of phase space regions that are only defined within the adiabatic approximation. It is only in the adiabatic approximation that the separatrix itself is a curve of zero measure; in reality there is a finite region around the separatrix within which adiabaticity breaks down. As Fig.~\ref{fig:decel}b) shows, indeed, the adiabatic breakdown around the separatrix means that orbits do not really remain within disjoint regions $A_{1\pm}$ but flow continuously through all three regions within a zone around the separatrices. The area of this adiabatic breakdown zone must also be considered within our Liouvillian logic, because this area too can change over time.

This issue may seem like a fatal flaw in the KNH formula as a prediction for post-adiabatic transition probabilities, but in fact the flaw can be remedied systematically, leading to an extended version of the KNH formula that really does work.

\section{III. Extending the KNH formula}

\subsection{A. Optimal canonical coordinates}
Adiabatic approximations break down near a separatrix because somewhere on the separatrix there is an unstable fixed point, where trajectories diverge and converge with infinite slowness, so that no finitely slow time dependence of the Hamiltonian can be slow in comparison to the evolution it generates \cite{Hannay}. It was therefore an important advance by Cary, Escande and Tennyson in 1986 to prove that one can always transform to canonical coordinates in which the unstable fixed point's location, and the quadratic Hamiltonian in its neighborhood, are exactly time-independent \cite{Tennyson}. This set of canonical coordinates thereby optimizes the accuracy of the adiabatic approximation.

As is usual for transformations that achieve things exactly, this optimal transformation can itself be hard to construct. One can often find a transformation, however, that will at least make the unstable fixed point change less---and this may improve the capture fraction estimate enough to make it a useful guide for control strategies. For our Hamiltonian (\ref{E:pertPendH}), for example, we can make the simple time-dependent canonical transformation $(\phi,P)\to (\phi,p)$ with $p=P-\alpha(t)$. The new Hamiltonian (that is, the ``Kamiltonian'' adjusted by adding the generator of the transformation \cite{Goldstein}) is then
\begin{equation}\label{Htrans}
	H_{2}=\frac{p^2}{2}+\dot\alpha\phi-\beta^{2}\cos(\phi).
\end{equation}

This new Hamiltonian $H_{2}$ is canonically equivalent to $H_{1}$ and hence generates the same exact evolution. If we are considering our system to represent a charged particle in an electric field, as described above, then the transformation from $H_{1}$ to $H_{2}$ is simply a gauge transformation such that the electric field is described in $H_{2}$ without any vector potential, using only the time-dependent scalar potential $V\von{\phi}=\dot\alpha\phi-\beta^2\cos\von{\phi}$. In this gauge it might appear natural to take $\phi \in \mathbb{R}$, instead of restricting $\phi$ to a ring as we were able to do in the original gauge. It is nevertheless possible to keep the restriction of $\phi$ to $\phi \in [-\pi,\pi[$, because $H_2\von{\phi+2\pi n,p,t}=H_2\von{\phi,p,t}+2\pi n \dot\alpha$ for every integer $n$. Hence shifting $\phi$ by $2\pi n$ only changes the instantaneous Hamiltonian by a constant, which has no effect on the equations of motion. We can therefore describe all orbits exactly while considering only the phase space for $-\pi \leq \phi < \pi$. Every portion of trajectory found in the region $-\pi+2\pi n \leq \phi < \pi+2\pi n$ has a counterpart in the region $-\pi \leq \phi < \pi$ that is exactly the same curve, just with a trivially shifted energy.

\begin{figure}[htb]
\centering
\includegraphics[width=.475\textwidth]{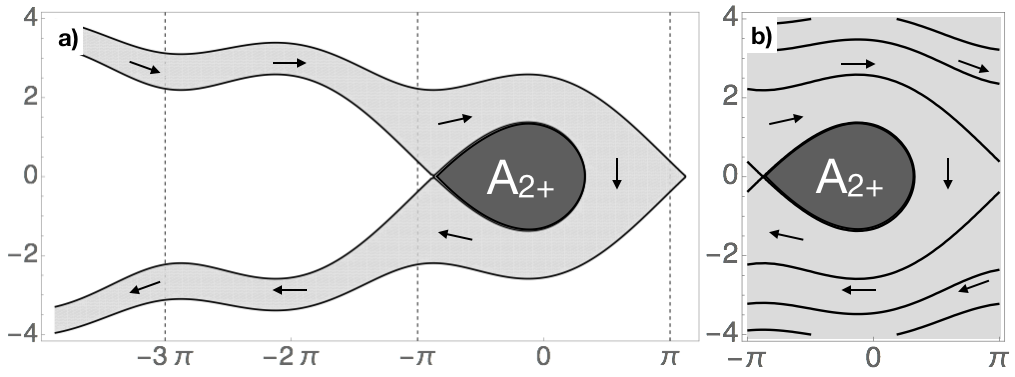}
\caption{\label{H2sep} The adiabatic separatrices of the transformed Hamiltonian $H_{2}$, plotted in the phase space of $\phi$ (horizontal axis) and $p$ (vertical axis), for the case $\dot{\alpha}=\sin(\pi/8)$ and $\beta=1$. The inner separatrix loop encloses the target region $A_{2+}$, shaded dark. Panel a) shows two successive separatrices in the extended range of $\phi$, with shading between them that gradually changes from gray to white, while panel b) shows the same separatrices and shaded region projected into the range $-\pi\leq\phi\leq\pi$. Adiabatic evolution of the system outside the $A_{2+}$ region is to flow between the separatrices without crossing them, as indicated by the arrows; post-adiabatic corrections can shift some orbits into $A_{2+}$ from outside it. In the projected picture of b), $\phi$ behaves as a periodic coordinate even though $H_{2}$ is not periodic in $\phi$.}
\end{figure}

The adiabatic approximation of $H_{2}$ is significantly different from that of $H_{1}$, however, as we can see from the instantaneous separatrix plots in the bottom panels of Fig.~\ref{fig:sep} and, enlarged and simplified for comparison with Figs.~\ref{fig:sepsketch} and \ref{H1sep}, in Fig.~\ref{H2sep}. One separatrix of $H_{2}$ is a closed loop that begins and ends at the fixed point (dark central loop in Figs.~\ref{H2sep}a and b); it therefore plays the role of both separatrices together under $H_{1}$. We can identify the interior of this closed separatrix as the `+' region for $H_{2}$, $A_{2+}$, since under the canonical transformation $p = P-\alpha$ it is mapped inside the $A_{1+}$ region of $H_{1}$. As the lower two panels of Fig.~\ref{fig:sep} show, the $H_{2}$ version of the target region $A_{2+}$ can indeed grow in time, even when $\beta$ is constant and so the $H_{1}$ version $A_{1+}$ remains constant as well.

The fact that $A_{2+}$ can be growing even when $A_{1+}$ is not is an encouraging sign that the KNH formula can perhaps be salvaged, and made to provide accurate estimates of capture or loss fractions after all, by changing to a canonical representation in which the adiabatic approximation is more accurate. This encouraging sign will turn out to be correct; for example the behavior of the $H_{2}$ separatrices will be able to explain both the non-zero capture and loss fractions that we have just shown as significant violations of the KNH formula. It is not quite enough just to transform from $H_{1}$ to the more adiabatic representation $H_{2}$, however, because as soon as we have transformed from $H_{1}$ to $H_{2}$ we find that we cannot even define the capture fraction $\mathcal{P}_{+}$ as we did in \ref{Pn0}.

Although the inner separatrix shown in Fig.~\ref{H2sep} lets us easily define the target region $A_{2+}$, we face a basic problem in trying to identify the two other regions $A_{20}$ and $A_{2-}$ that are logically necessary for the KNH argument.  The other separatrices of $H_{2}$ are \emph{open}. If the range of $\phi$ is allowed to be infinite then these separatrices run infinitely away to the upper and lower left; if we project into the region $\phi\in[-\pi,\pi]$ as discussed above, then as shown in Fig.~\ref{H2sep}b) they wrap around and around through $\phi$ while running indefinitely to higher and lower values of $p$. The $H_{2}$-separatrices thus divide the phase space with periodic $\phi$ into only two separate regions, not three as we had with $H_{1}$---namely $A_{2+}$ and the infinitely wrapping corridor that flows around $A_{2+}$. If we try to assign the dark, light gray, and white shadings of Figs.~\ref{fig:sepsketch} and \ref{H1sep} now in Fig.~\ref{H2sep}, the wrapping corridor can only have shading that changes gradually from gray to white. Its upper and lower ends must correspond somehow to the $A_{10}$ and $A_{1-}$ of the $H_{1}$ representation, but there is no longer any obvious boundary to divide the wrapping corridor of $H_{2}$ into two such regions.

The logic of the KNH formula was based on having three regions---a donor region $A_{0}$ and two recipients $A_{\pm}$---with the system choosing between $A_+$ and $A_-$ when it leaves $A_{0}$. We therefore cannot apply the KNH capture formula \eqref{Pn0} as it stands to cases like that in $H_{2}$ where there are only two adiabatic regions. We can however apply the same Liouvillian logic that led to the KNH formula to derive an extended KNH formula that \emph{can} be applied to Hamiltonians like our $H_{2}$.

\subsection{B. Capture fraction with an open separatrix}
This way of extending the KNH formula was indicated in 1994 by Chernikov and Schmidt \cite{Chernikov}, in a paper on adiabatic chaos in Josephson junction arrays. Our derivation of this extension of KNH will be motivated by Section III.C of Ref.~\cite{Chernikov}, with some generalization; in particular Chernikov's and Schmidt's Fig.~8 may be compared directly to our Fig.~\ref{H2sep} and others. In this subsection we will present an abbreviated sketch of the derivation, leaving the full details for the Appendix. 

\begin{figure}[htb]
\centering
\includegraphics[width=.475\textwidth]{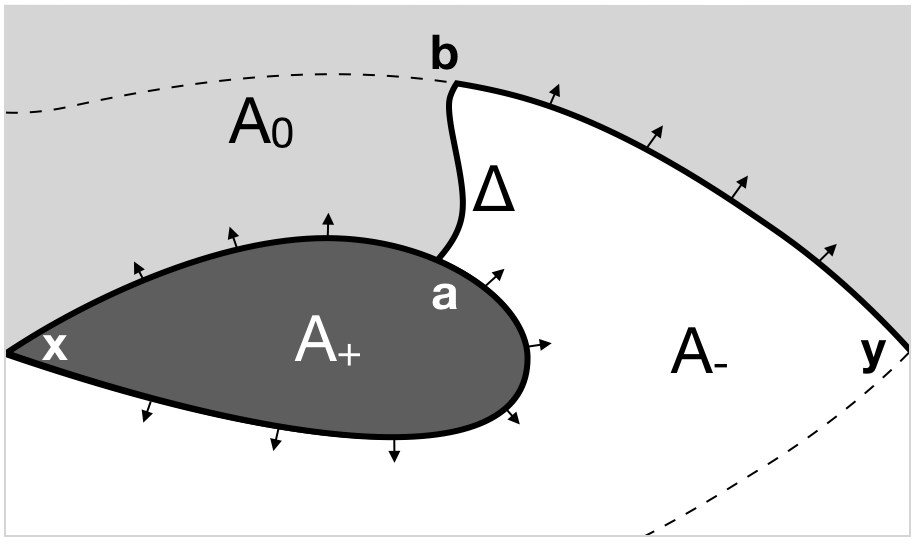}
\caption{\label{Fig6} Contours and regions relevant to the extension of the KNH formula. The thickly drawn curves include the entire inner separatrix and part of the outer separatrix (the remainder of the outer separatrix being shown in thinner dashed lines). Small arrows indicate that these separatrices are both slowly deforming in time (in this case, expanding). Points $\mathbf{x}$ and $\mathbf{y}$, at the left and right edges of the plot, are unstable fixed points. The points $\mathbf{a}$ and $\mathbf{b}$ are arbitrarily chosen points on the two separatrices, while the curve $\Delta$ is an arbitrary curve connecting $\mathbf{a}$ and $\mathbf{b}$. The addition of the arbitrarily constructed curve $\Delta$ completes the division of phase space into three regions: $A_{+}$ (dark), $A_{0}$ (light gray), and $A_{-}$ (white).}
\end{figure}

Our abbreviated derivation will be based on the phase space sketch Fig.~\ref{Fig6}, which shows a region around the closed inner separatrix of a generic Hamiltonian like that shown in Fig.~\ref{H2sep}. The labelled contours and points in Fig.~\ref{Fig6} are related to adiabatic energy contours. The full derivation in the Appendix also uses some slightly different points and contours, defined in terms of exact trajectories, because the main work of the full derivation will be to show how exact quantities can be approximated accurately in terms of adiabatic ones. The abbreviation which we make here is to gloss over the distinction between exact and adiabatic trajectories and mention only the adiabatic ones.

The crucial step in constructing the extended KNH formula is to introduce the curve $\Delta$ as an arbitrarily drawn curve which connects an arbitrarily chosen point $\mathbf{a}$ on the closed separatrix with another arbitrarily chosen point $\mathbf{b}$ on the open separatrix, and thus provides an artificial division between the regions  $A_{0}$ and $A_{-}$. The KNH logic about incompressible phase space flow from $A_{0}$ into both $A_{+}$ and $A_{-}$ can now be applied as it was before, when all boundaries between the regions were adiabatic separatrices. Under time evolution now, however, system points will flow through the non-separatrix artificial border ${\Delta}$, even within the adiabatic approximation. 

At any time $t$ we can nonetheless still adiabatically compute the total flux of phase space out of $A_0\von{t}$. One term in this flux is simply the area shrinkage rate $-\dot A_0\von{t}$ that we considered before. To this we must now simply add the flux $\Phi_-\von{t}$ of points evolving into $A_{-}$ across the curve $\Delta$. For the fraction of orbits which exit $A_{0}(t)$ at time $t$, and are captured into $A_+\von{t}$, we therefore obtain
\begin{eqnarray}\label{P+a}
	\mathcal P_{+}(t) = \frac{\dot{A}_{+}(t)}{\Phi_{-}(t)-\dot{A}_{0}(t)}.
\end{eqnarray}
The rates of change of the areas can be computed as before---and since ${\Delta}$ is arbitrary we are free to choose it in the way that will make the areas easiest to compute. The flux through ${\Delta}$ is even easier to compute, since it is an identity of Hamiltonian time evolution (see the Appendix) that the instantaneous flux through any curve in phase space is equal to the difference between the instantaneous values of the Hamiltonian at the curve's endpoints. The endpoints $\mathbf{a}$ and $\mathbf{b}$ of the curve ${\Delta}$ lie on separatrices, and since separatrices are contours of constant energy, the values of the Hamiltonian at the endpoints $\mathbf{a}$ and $\mathbf{b}$ of ${\Delta}$ are the same as the values of the Hamiltonian at the adiabatic fixed points $x$ and $y$, respectively. We can therefore express our extended KNH formula (\ref{P+a}) more explicitly as 
\begin{eqnarray}\label{P+0text}
	\mathcal P_{+}(t) = \frac{\dot{A}_{+}(t)}{H\von{{\mathbf{x}},t}-H\von{{\mathbf{y}},t}-\dot{A}_{0}(t)}.
\end{eqnarray}

One might well be concerned that this revised formula for $\mathcal{P}_{+}$ now depends on the arbitrary curve ${\Delta}$, since the area $A_{0}(t)$ depends on how the artificial part of the border of $A_{0}$ is drawn. Indeed $\mathcal{P}_{+}$ does depend on the precise choice of ${\Delta}$, but---as we explain in the Appendix---not to leading order in the small adiabatic slowness parameter. The KNH formula was never more than leading-order adiabatic approximation anyway and so the sub-leading dependence of the extended formula on the arbitrary ${\Delta}$ does not matter. If the time-dependence of the Hamiltonian is slow enough for this whole approach to capture fraction estimation to be valid, choosing a different ${\Delta}$ will make only tiny changes in $\mathcal{P}_{+}$, comparable in size to the higher order post-adiabatic corrections which are present in any case.

In (\ref{P+0text}) we have thus found a natural adjustment of the KNH formula to cases where there is adiabatic flow between two of the three regions, as well as slow deformation of separatrix borders. The formula is still simple enough to be a useful guide for control strategy, inasmuch as it provides an estimate for the probability of the dynamical transition which can be obtained directly from the instantaneous Hamiltonian, without having to solve for any time evolution. The basis of the KNH formula in Liouville's theorem is moreover intact; the total amount of incompressible phase space which moves out of $A_{0}$ has simply been recognized to include the non-zero flux through ${\Delta}$.
The merit of this extended formula is that it can still be applied when canonical transformations that make fixed points less time-dependent, and thereby improve the accuracy of adiabatic theory, have somehow removed one of the separatrix borders between two of the three KNH regions.

\subsection{C. Example of application of the extended KNH formula}

\begin{figure}[htb]
\centering
\includegraphics*[trim=40 20 45 70, width=.475\textwidth]{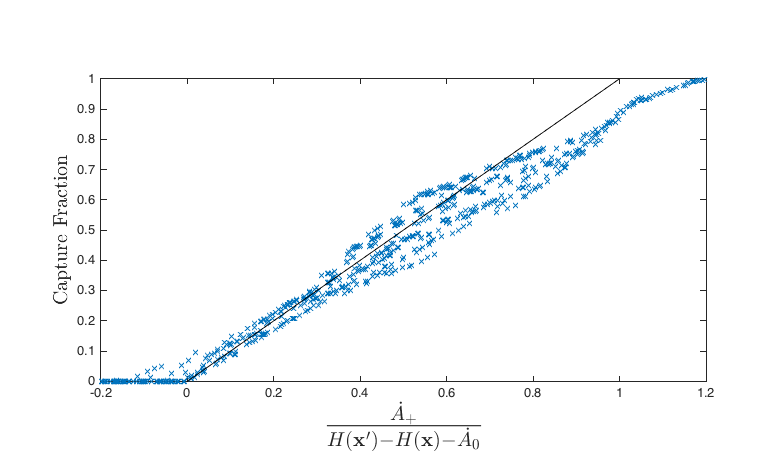}
\caption{\label{fig:EKNH} \textit{Accuracy of the extended KNH formula (\ref{P+0text})}. The crosses are values for the capture fractions obtained from numerically evolved ensembles as explained in the text. The diagonal black line represents equality between the adiabatically estimated and numerically exact capture fractions.}
\end{figure}

When the principle of Ref.~\cite{Tennyson} is applied to maximize the accuracy of the adiabatic approximation by making fixed points immobile (or even just nearly so), and when the extended KNH formula (\ref{P+0text}) is adopted as needed, the basic idea of using Liouville's theorem to deduce probabilities of dynamical transitions is thus confirmed as robustly general. As one illustration we show in Fig.~\ref{fig:EKNH} a plot of numerically exact capture fractions for our $H_{1}$/$H_{2}$ model (the exact evolutions being identical for the two canonical representations), versus the adiabatic prediction of (\ref{P+0text}), for a set of many different time dependences of the parameters.

In particular we take a `pre-initial' ensemble at time $t=0$ of 1000 different phase space points. In this ensemble the initial values of the angle $\phi$ are uniformly spread over $\ebraces{-\pi,\pi}$ and the energy (which determines $p$) is uniformly spread over $\ebraces{99,101}$. To prepare the initial ensemble which will be used to test the extended KNH formula, we then evolve these `pre-initial' points numerically under the Hamiltonian \eqref{Htrans} with the constant parameters $\dot\alpha\von{t}=0.01$ and $\beta\von{t}=1$, until the first points of our pre-initial ensemble arrive at the separatrix under $H_{1}$. (This occurs at $t=t_{s}\doteq 1287$.)

From this time $t_{s}$ onwards we compare 1500 different cases of time-dependent Hamiltonians; for each of the 1500 cases we follow the evolution of our entire 1000-trajectory ensemble. The different cases of time dependence of our parameters are that $\dot\alpha\von{t}=0.01 + \varepsilon_\alpha t$ and $\beta\von{t}=1+\varepsilon_\beta t$, with uniformly distributed $\varepsilon_\alpha \in \ebraces{-0.5,1} \times 10^{-3}$ and $\varepsilon_\beta \in \ebraces{-7,7} \times 10^{-3}$. These ranges of $\varepsilon_{\alpha,\beta}$ were chosen because the edges of the ranges were estimated to give capture fractions greater than zero or less than one. For each of these 1500 cases of parameter time dependence, we continued the Hamiltonian evolutions of our 1000 trajectories and noted what fraction of them were captured into region $A_{2+}$ as defined under the Hamiltonian $H_{2}$. Every blue cross in Fig.~\ref{fig:EKNH} denotes this capture fraction for one of the 1500 $\varepsilon_\alpha,\varepsilon_\beta$ pairs, plotted versus the capture fraction estimated for its case of parameter time dependence according to the extended KNH formula \eqref{P+0text}, evaluated at the time $t_{s}$, using $A_{2\pm,0}$ in the roles of $A_{\pm,0}$.

Fig.~\ref{fig:EKNH} confirms that the extended KNH formula works excellently in most cases, and quite well in all cases, as long as the capture fraction is not too high. The spread of numerical points around the analytical line represents the inevitable limitation of the adiabatic approximation, including the fact that not all points in our trajectory ensemble actually meet the separatrix at the same time $t_{s}$. 

At larger capture fractions the post-adiabatic scatter increases and the numerical trend also falls below the analytical estimate systematically. The extended KNH formula still remains good enough, however, that it can explain its own comparative weakness in this regime: the higher capture fractions are reached because $A_{2+}$ is changing more rapidly, but the more rapid time dependence of the Hamiltonian means that the adiabatic approximation, on which the capture fraction estimate is based, becomes less accurate. 

In fact the discrepancy shown in Fig.~\ref{fig:EKNH} between the extended KNH formula and the numerical capture fraction appears systematic enough that one can anticipate being able to improve the estimate with some systematic post-adiabatic corrections, especially since for larger capture fractions the post-adiabatic scatter decreases as well. It is not clear to us, however, whether such a further improved KNH formula would really be worth using: the most accurate estimate of all can always in principle be found by numerically evolving a large ensemble of trajectories, exactly as we did to prepare Fig.~\ref{fig:EKNH}, and the value of analytical estimates like the extended KNH formula lies only in the fact that they are much easier than that to compute. Unless it remained quite simple, a more accurate analytical estimate might be self-defeating. 

In any case we leave this possibility of further improvement to the KNH formula for future work, and conclude this Section of our paper with the confirmation that even though it retains a simple form in \eqref{P+0text}, it does work very well, as long as the time dependence of the Hamiltonian is slow enough for adiabatic methods of any kind to apply. Once the more significant failures of adiabaticity due to a moving unstable fixed point are removed, by transforming to a canonical representation in which the unstable fixed point moves only slightly, the derivation of the extended KNH formula  \eqref{P+0text} that we give in our Appendix becomes valid, and the potential contradictions involved in predicted non-adiabatic evolution based on adiabatically defined areas are avoided. Simple reasoning about incompressible phase space flow and separatrix growth really can provide accurate estimates for how likely it is that a given orbit will undergo a separatrix-crossing dynamical transition, without having to actually solve the non-adiabatic time evolution.

\section{IV. Control applications of Liouville's theorem}
Having shown that the Liouvillian picture of dynamical transitions can be accurate when correctly applied, we will now present some examples to show how it can be the basis for control techniques. 

\subsection{A. A transportation task}
Following \cite{bucket,Bazzani} we first consider a new control task: instead of capturing the system into $A_{+}$, our goal now will be to keep the system inside $A_{+}$ in order to transport the system in phase space. If the whole $A_{+}$ region steadily moves though phase space, then orbits retained within it adiabatically are carried along like so much water in a bucket. In particular \cite{Bazzani} considers our same model system with $H_{1}$ from (\ref{E:pertPendH}), with $\beta$ held constant, but with $\alpha$ rising linearly with time in order to make $A_{1+}$ likewise rise steadily in the $\phi,P$ phase space plane and (hopefully) carry system orbits along with it from low $P$ to high $P$.

\subsubsection{Empirical law for transport losses}
\begin{figure}[htb]
\centering
\includegraphics*[trim=210 140 180 90, width=.475\textwidth]{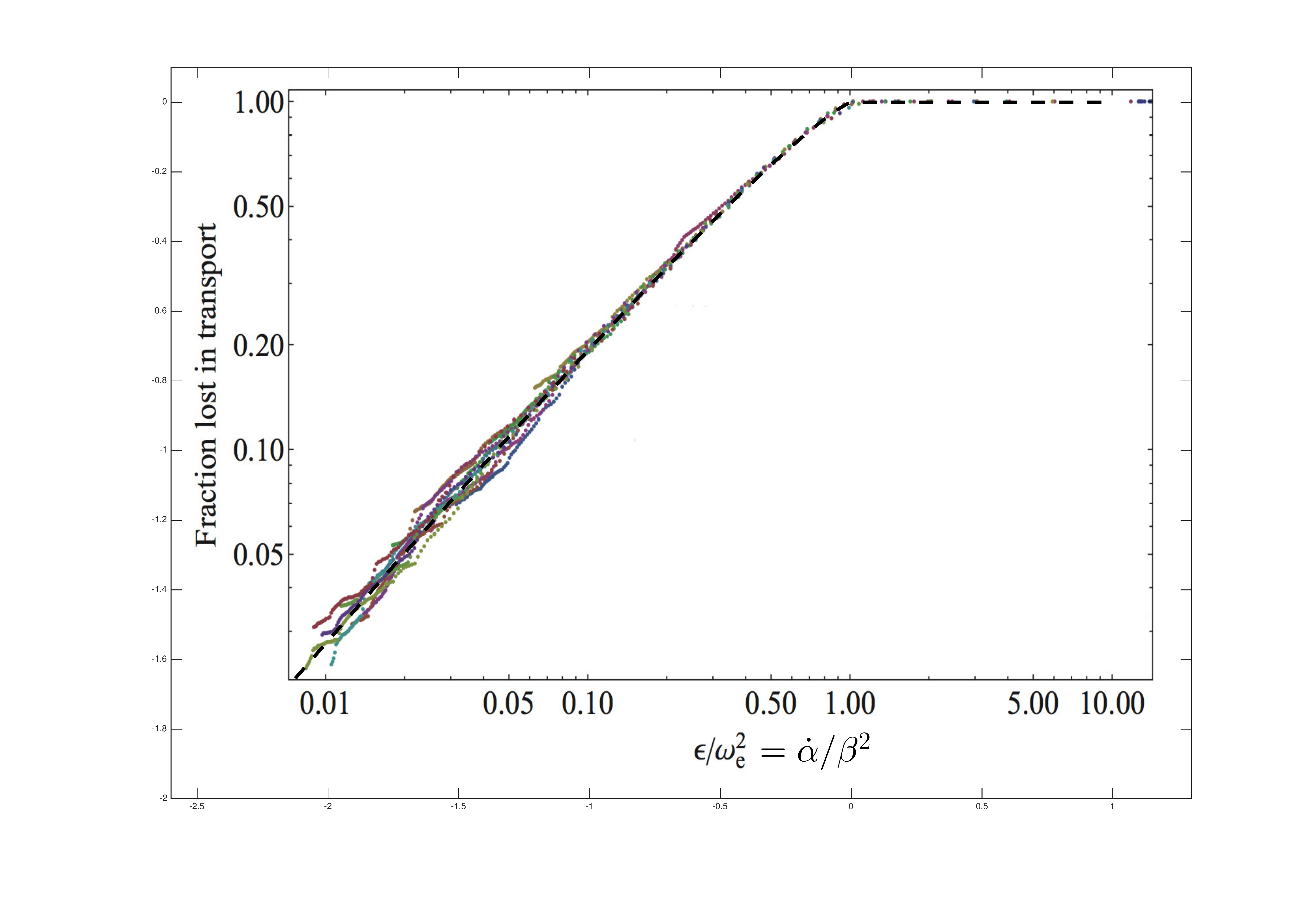}
\caption{\label{fig:overlay} {\bf Colored dots:} From \cite{Bazzani} by Bazzani \textit{et al.}, the `spillage' fraction $\nu$ as defined in the text versus the quantity represented by us as $\dot{\alpha}/\beta^{2}$ and by Bazzani \textit{et al.} as $\varepsilon/\omega_{e}^{2}$. Note that the horizontal axis is plotted logarithmically. Differently colored points refer to different values of $\beta$ and $\dot{\alpha}$. The diagonal pattern of dots is not quite a straight line, but its lower left half is fit very closely by the numerical law \ref{Bazzanilaw} that was reported in \cite{Bazzani}. {\bf Dashed black contour:} Calculated value $1-A_{2+}/A_{1+}$. This result is in fact exact for $\nu$, because $H_{2}$ is time-independent, and so all the colored points should lie exactly along this slightly bending dashed line. The scatter of the points from \cite{Bazzani} around this curve must represent sampling or other numerical errors.}
\end{figure}

As already noted in Section II above, however, Ref.~\cite{Bazzani} by Bazzani \emph{et al.} reports that this procedure is not perfectly efficient. If an initial ensemble of orbits fills the entire $A_{1+}$ region defined by $H_{1}$, numerical evolution had some orbits escape almost immediately: instead of being carried, they spill out of the bucket. Ref.~\cite{Bazzani} evolved this initial set of orbits for a large range of parameter time dependencies, and for each such case of time dependence, computed the fraction $\nu$ of initial points that were thus `spilled' instead of being transported. The authors of Ref.~\cite{Bazzani} were then able to fit the numerically obtained transport fraction $\nu$ with a numerically empirical formula which in our notation reads
\begin{eqnarray}\label{Bazzanilaw}
	\nu=1.132\left(\frac{\dot{\alpha}}{\beta^{2}}\right)^{0.754}\;.
\end{eqnarray}
The numerical data on which this empirical law was based are reproduced from Ref.~\cite{Bazzani} in Fig.~\ref{fig:overlay}, together with a dashed curve which is actually \emph{not} the empirical formula (\ref{Bazzanilaw}) from \cite{Bazzani}, but rather the exact curve which we will derive here below.

\subsubsection{Exact transport loss}
\begin{figure}[htb]
\centering
\includegraphics*[trim=80 10 100 20, width=.475\textwidth]{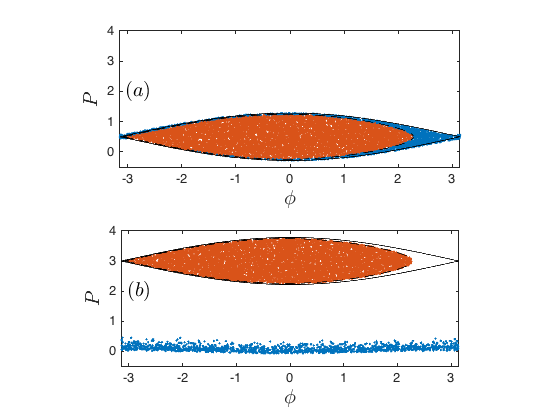}
\caption{\label{fig:transport} \textit{Survival fraction}. Curves are separatrices as in Fig.~\ref{fig:decel}: solid are the separatrices of $H_{1}$, dashed the inner separatrix of $H_{2}$. In (a) 20 000 initial points are uniformly distributed at $t=0$ inside the $H_{1}$-separatrix for $\alpha = 0.5$, $\beta=\pi/8$. In (b) about 14\% of the initial points have spilled out of the separatrix region $A_{1+}$ after evolution with $\beta=\pi/8$, $\alpha = 0.5 + t/100$ until $\alpha=3$. Those points that will remain inside $A_{1+}$ in (b) are shown in red in both (a) and (b), while the spilled points are blue. We see in (a) that the successfully transported points are precisely those that are inside the $H_{2}$-separatrix, in region $A_{2+}$.}
\end{figure}

In Ref.~\cite{Bazzani} it is explicitly noted that the Hamiltonian $H_{1}$ can be transformed into $H_{2}$; that this defines a new separatrix inside the separatrix defined by $H_{1}$; and that one can expect all orbits initially within this $H_2$-separatrix to be retained and transported while those outside the $H_2$-separatrix will immediately be `spilled'. And indeed this is precisely what happens: see Fig.~\ref{fig:transport}.

Through an apparent oversight on this one narrow point within a lengthy paper that substantially advanced the whole control perspective on adiabatic dynamical transitions, Ref.~\cite{Bazzani} explicitly recognizes the importance of the $H_{2}$-separatrix, and yet only ``suggests'' qualitative explanations of which the empirical scaling law (\ref{Bazzanilaw}) ``could be a consequence''.  In fact the entire initial volume of the $H_{2}$-separatrix \textit{must} remain within the $H_{2}$ separatrix forever, and thus be successfully transported in $P=p-\alpha(t)$, because with $\beta$ constant and $\alpha$ linear in $t$, $H_{2}$ is time-independent. The $H_{2}$-separatrix is in this special case not merely an adiabatic separatrix, but an exact one. All points that are initially outside the $H_{2}$-separatrix are correspondingly `spilled'.

\begin{figure}[htb]
\centering
\includegraphics*[width=.475\textwidth]{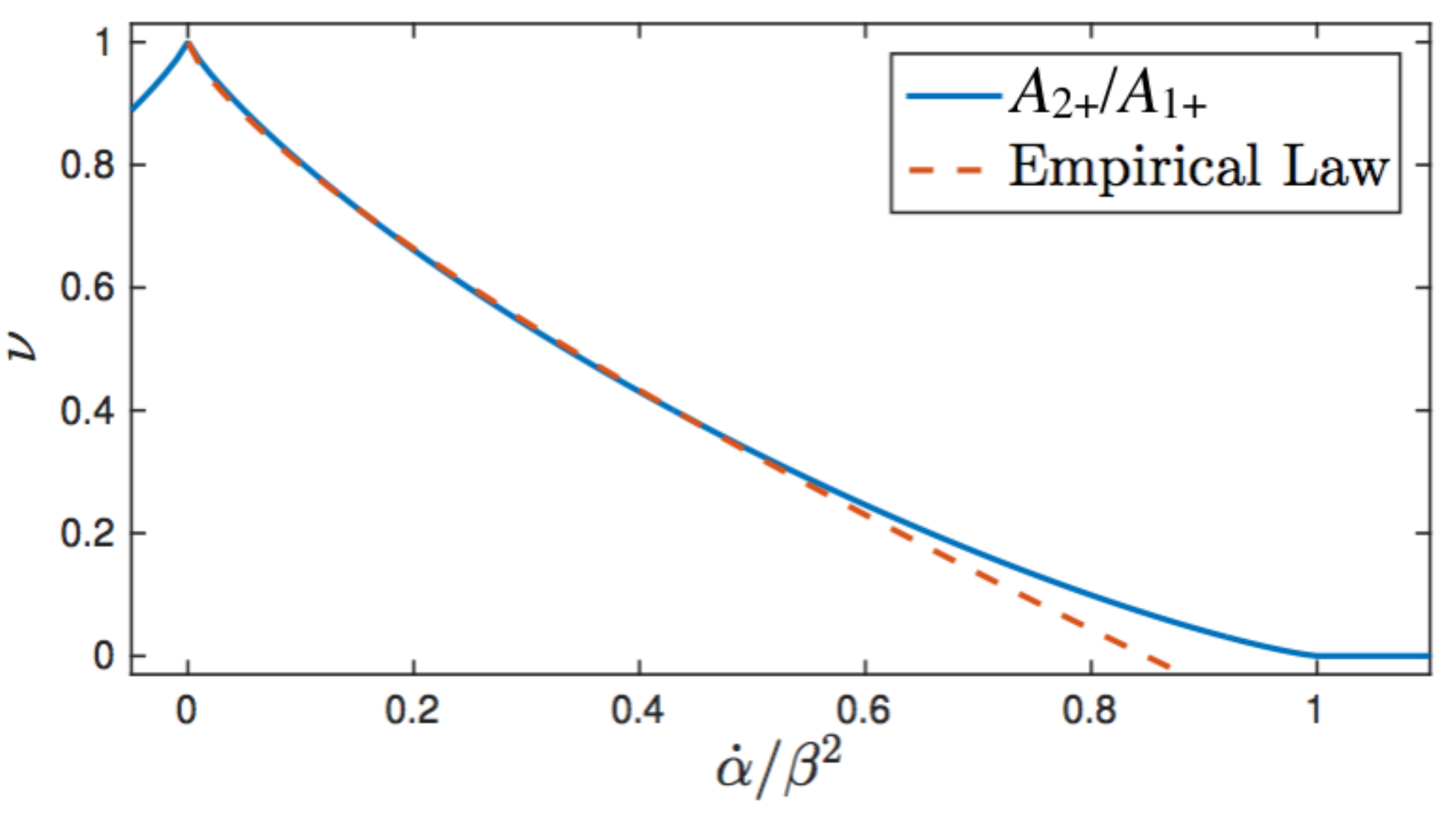}
\caption{\label{fig:areafraction} {\bf Solid blue line:} The fraction $\nu=A_{2+}/A_{1+}$ versus $\dot{\alpha}/\beta^2$. {\bf Red dashed line:} Empirical law $\nu=1-1.132\gamma^{0.754}$ published in \cite{Bazzani}.}
\end{figure}

The $H_{2}$ separatrix can be determined analytically and its area can be computed numerically. Since the initial ensemble of \cite{Bazzani} is a uniform filling of the $H_{1}$-separatrix, whose area we computed analytically in Eqn.(\ref{separea}), we can easily obtain the exact $\nu$ as $A_{2+}$ divided by $A_{1+}$. The results are shown in Figs.~\ref{fig:overlay} and \ref{fig:areafraction} to reproduce the extensive numerical simulations of Bazzani \emph{et al.} very well. The fact that the area results are exact in this case is a special feature of the exactly time-independent $H_{2}$, but the general principle from Ref.~\cite{Tennyson} of using coordinates in which fixed points stay fixed tells us that we can expect good accuracy from the area-based estimates whenever $H_{2}$ depends slowly on time.

\subsubsection{A bigger bucket}
The initial ensemble filling $A_{1+}$ uniformly was achieved in \cite{Bazzani} as the product of a pre-initial stage of adiabatic capture from a simpler ensemble, by raising $\beta$ slowly from 0. In contrast to the earlier literature's focus on solving initial value problems, the two-stage process of capture and transport was explicitly conceived in \cite{Bazzani} as a control protocol, and the fact that orbits were invariably lost in the second stage of transport was interpreted as suggesting that such two-stage strategies might be sub-optimal in general because the high capture efficiency of the first stage could be outweighed by the mediocre transport efficiency of the second.

\begin{figure}[thb]
\centering
\includegraphics*[trim=80 10 100 20, width=.475\textwidth]{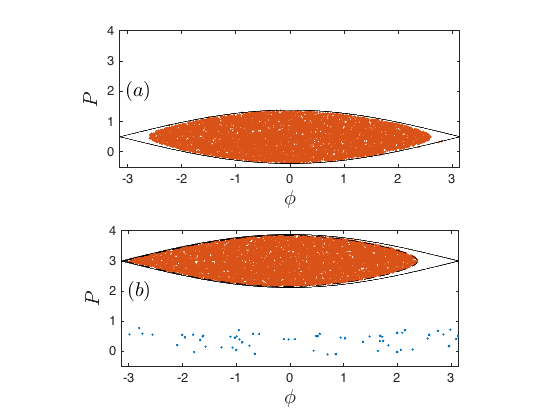}
\caption{\label{fig:lossless} Before the transportation process is started the area inside the separatrix is increased such that $A_{2+}$ slightly exceeds the measure of the initial ensemble of orbits. Afterwards the transportation process is started by smoothly accelerating to the target speed $\dot{\alpha}=10^{-2}$. Almost all points are successfully transported.}
\end{figure}

A Liouvillian perspective based on the incompressibility of phase space, however, suggests a simple remedy for the second-stage loss problem. Don't try to move a bucket that is full to the brim; if the task is to transport a given measure of phase space, use a bigger bucket that will not be so full. We can implement this idea for the same initial ensemble shown in Fig.~\ref{fig:transport}(a) by not transporting yet right away at $t=0$, but instead slowly raising $\beta$ further until $A_{2+}$ will be large enough to contain the entire ensemble, and only then beginning the transport stage of the two-stage protocol. In Fig.~\ref{fig:lossless} we show the results of a procedure in which $\beta$ is thus raised prior to transport from $\pi/8\doteq 0.39$ to $1/\sqrt{5}\doteq 0.45$. At this higher value of $\beta$ the area $A_{2+}\doteq 6.4$ is slightly greater than the measure of the initial ensemble, which is $2\pi$.

When we do begin transport it is another simple-minded improvement to avoid suddenly jerking the bucket, but rather accelerate it gradually. The procedure shown in Fig.~\ref{fig:lossless} thus also lets $\alpha$ grow quadratically as $\alpha(t<100)=0.5 \braces{1+10^{-4}t^{2}}$ until $t=100$ and only thereafter grow linearly as $\alpha(t>100)=1+10^{-2}(t-100)$. By thus gently moving a larger bucket, Fig.~\ref{fig:lossless} shows that we can achieve over 99\% transport efficiency. Our point is not that the reasoning behind this scheme is non-trivial, but precisely that it is simple enough to be applied quite robustly even in more complex systems.

\subsection{B. Capture through Liouville control}
\subsubsection{Designing Hamiltonians for control}
The model $H_{1}$ of \eqref{E:pertPendH} is a much-studied paradigm system in adiabatic theory. It has two time-dependent control parameters $\alpha(t)$ and $\beta(t)$, and as in Ref.~\cite{Bazzani} one may compare alternative protocols for their time dependence to see which most efficiently achieves a given goal. With developments in nanotechnology and in highly controllable experimental systems such as quantum gases, however, it has become possible to consider not only system parameters to be chosen at will, but even the functional form of Hamiltonian terms. With increased understanding of molecular machinery in biological systems, furthermore, we may one day be able to understand how different kinds of molecular mechanism produce chemical behavior that effectively controls these complex machines. It is therefore worthwhile to compare alternative control protocols that differ not only in the time dependence of parameters but in the functional form of coupling terms through which control is applied.

As a simple example of this we will compare the following two Hamiltonians:
\begin{eqnarray}\label{}
	H_{A}&=&\frac{P^{2}}{2}-\alpha(t) P - c \sqrt{P}\cos\phi\nonumber\\
	H_{B}&=&\frac{P^{2}}{2}-\alpha(t) P - c \sqrt{P(1-P)}\cos\phi\;,
\end{eqnarray}
where $c$ is a constant (which can be made the same for $H_{A}$ as for $H_{B}$ without loss of generality by rescaling $P$, $\alpha$ and $t$ in $H_{A}$). For both Hamiltonians we require $P\geq 0$ and for $H_{B}$ we must also have $P\leq 1$. 

In effect both these models $H_{A,B}$ are much like $H_{1}$, except with $\beta(t)$ now made into a dynamical variable rather than an external parameter. In reality, all time-dependent parameters in any model are dynamical variables, if we extend our dynamical description to include whatever apparatus is imposing their time dependence, and so comparing alternative control protocols actually is comparing different forms of Hamiltonian, in any case. We are now simply doing this explicitly. 

The whole phenomenology of our $H_{1}$ model is repeated in both these new models, including the time-dependent separatrices. We will consider cases where $c$ is quite small, so that it is easy to see that the separatrices for both $H_{A,B}$ will lie within a small range of $P$ around $P=\alpha(t)$, as long as $\alpha>0$ (and $\alpha<1$ for $H_{B}$). We can therefore expect that $H_{A}$ will be much like a case of $H_{1}$ with $\beta^{2}(t)\propto\sqrt{\alpha(t)}$, while $H_{B}$ should resemble a case of $H_{1}$ with $\beta^{2}(t)\propto\sqrt{\alpha(t)(1-\alpha(t))}$. With that in mind, all our Liouvillian KNH results and our understanding of $H_{1}$ separatrices should allow us to anticipate the performance of both $H_{A,B}$ at least qualitatively, without solving any actual equations of motion. The reader may wish to pause and predict before reading ahead: what will happen in each case if we try to use a decreasing $\alpha(t)$ to transport orbits downward in $P$? (In this case it will not even be necessary to transform to any new Hamiltonians analogous to $H_{2}$; the naive KNH formula (\ref{Pn0}) without the extension (\ref{P+0text}) will suffice to predict the drastically different behaviors in these two cases.)

\subsection{Two alternatives compared}
\begin{figure*}[htb]
	\centering
	\includegraphics*[trim=40 95 50 100 ,width=1\textwidth]{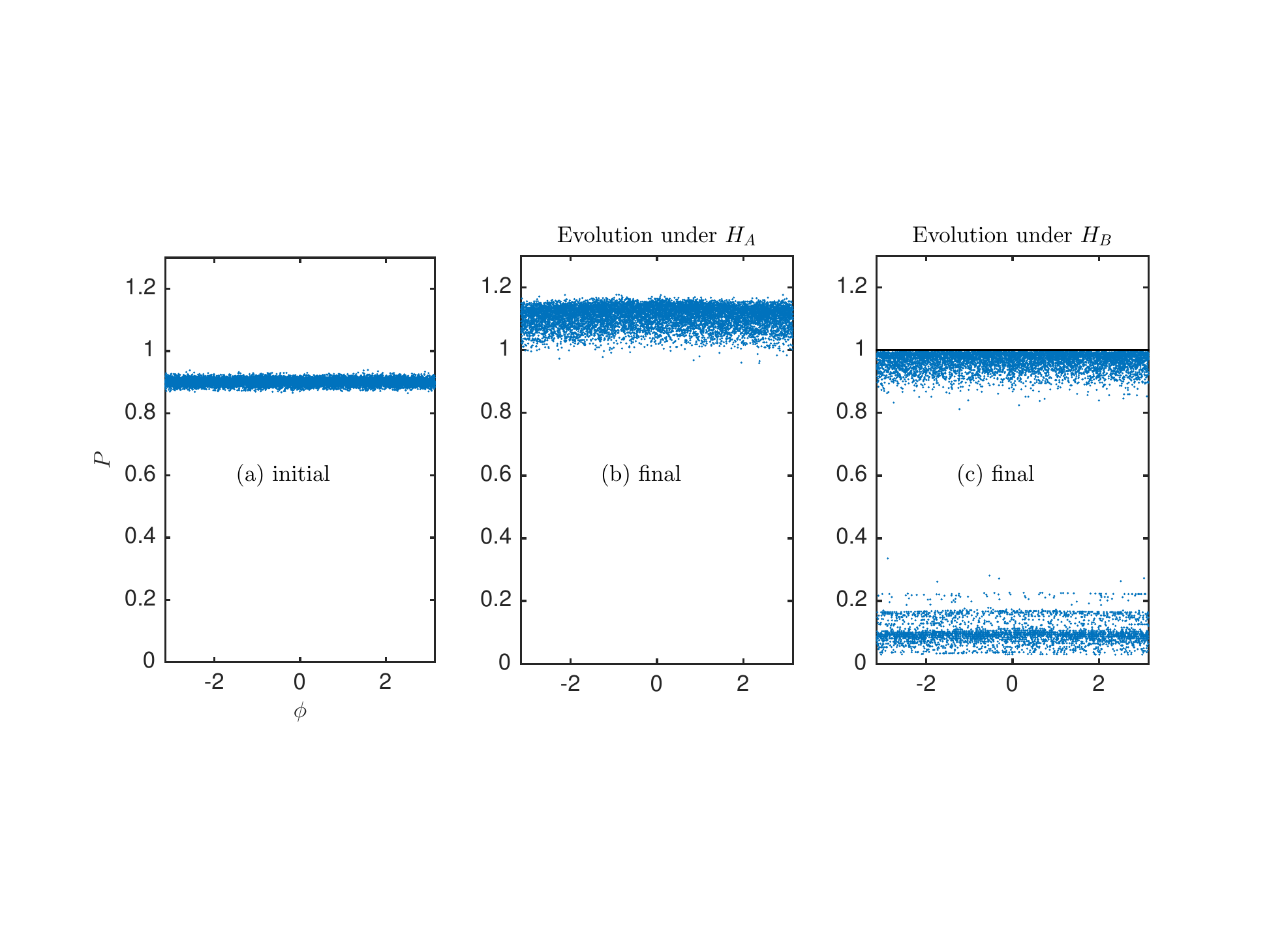}
	\caption{\label{fig:initialfinal} The $(\phi,P)$ phase space distribution of 10 000 orbits at the very early initial time (a), and the very late final time after evolution under $H_{A}$ (b) and $H_{B}$ (c), for the coupling constant $c=10^{-2}$ in both cases. At the initial and final times $\alpha(t)$ is so large (whether positive or negative) that orbits under both Hamiltonians are essentially lines of constant $P$.}
\end{figure*}

With only $\alpha(t)$ now left as a time-dependent parameter, we can still ask how efficiently we can capture and transport system orbits with the separatrices that $H_{A}$ and $H_{B}$ in general both have. As a concrete example consider an initial ensemble of orbits in which $\phi$ is distributed uniformly around the full circle $[0,2\pi)$ while $P$ is distributed with a narrow Gaussian weight around the mean $\bar{P}=0.9$. We assume that the task is to capture and transport these initial orbits to lower $P\sim 0.1$, and that to do this one can use either $H_{A}$ or $H_{B}$ with any $\alpha(t)$ one may wish. The results of two seemingly reasonable protocols, one for each Hamiltonian, are shown in Fig.~\ref{fig:initialfinal}.

Both protocols provide a separatrix that slowly moves down from $P=1$ to $P=0$. This means that for both Hamiltonians there exist orbits which remain inside the separatrix as it moves and therefore fulfil our control goal. Naively, indeed, one might expect both protocols to be workable, because they both clearly do tend to move orbits down from high $P$ to low when $\alpha(t)$ is decreasing. Nevertheless only the protocol with $H_{B}$ succeeds in transporting orbits to the target region, which it achieves for about 36\% of the initial orbits. Using $H_{A}$, not only do we fail to capture any of the initial orbits into the separatrix, and therefore fail to bring any orbits down to low $P$. In fact we make matters worse with $H_{A}$, in the sense that we displace all our initial orbits to even higher values of $P$.

\subsubsection{The Liouvillian explanation}
\begin{figure*}[htb]
	\centering
	\includegraphics*[trim=80 80 90 60 ,width=1\textwidth]{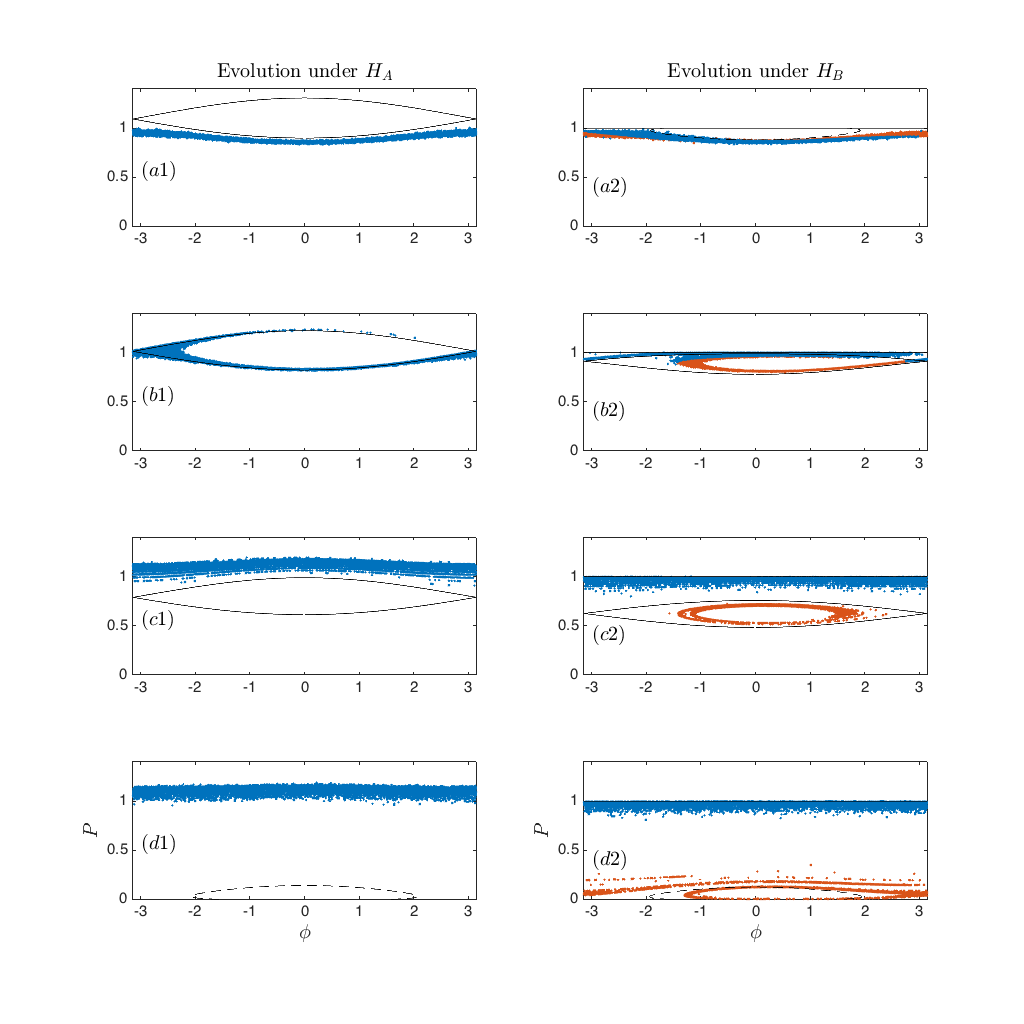}
	\caption{\label{fig:twoseps} The same evolutions shown in Fig.~\ref{fig:initialfinal} under the Hamiltonians $H_{A}$ (left coloumn) and $H_{B}$ (right coloumn), together with the instantaneous separatrices (solid black curves) and the orbits that will transform into the separatrix (dashed black curve). Four successive intermediate times are shown (top to bottom), to reveal how the two control schemes respectively fail and succeed.}
\end{figure*}

The reason for the failure of $H_{A}$ and the success of $H_{B}$ is clear from the Liouvillian perspective of the KNH formula, even without doing any difficult calculations at all. If $P$ is decreasing in the vicinity of $P=1$, then the $P$-dependent prefactors of the $\cos\phi$ terms in $H_{A}$ and $H_{B}$ are respectively shrinking and growing. The $\Sigma_{+}$ separatrix is therefore expanding as it moves down under $H_{B}$, but shrinking under $H_{A}$. Both these statements are easily deduced by inspecting $H_{A}$ and $H_{B}$, but they may be confirmed by looking at the separatrices in Fig.~\ref{fig:twoseps}, which shows the same evolutions whose very early and very late states were shown in Fig.~\ref{fig:initialfinal}, but for a series of four intermediate times which show how capture and transport either occur or fail to occur. 

There are many orbits of $H_{A}$ which will be transported down in $P$ as desired, but they are all orbits which are \emph{already} inside the separatrix, and transporting down, when the separatrix meets our initial ensemble. Because phase space is incompressible, there is no room for new orbits to enter the $H_{A}$ separatrix from outside; the KNH capture probability vanishes. We can even use Liouville's theorem to anticipate the otherwise surprising fact that the evolution under $H_{A}$ systematically displaces all the initial orbits upward in $P$. The shrinking separatrix is bringing new phase space down from high $P$ to low, and shedding orbits as it shrinks. Since phase space flows incompressibly, orbits that were initially present at lower $P$ before the separatrix descended through them must all move upwards to make room for the newcomers.

Under $H_{B}$, in contrast, the $\Sigma_{+}$ separatrix automatically grows as it moves down from $P\sim 1$. The KNH capture probability is significant (about 36\% in the example shown in Fig.~\ref{fig:initialfinal}), because the incompressibility of Liouvillian flow means that orbits \textit{must} be drawn into the separatrix. It inhales them like a lung drawing air.

In the previous Section III we showed how the simple KNH formula of Section II could be improved into a robustly accurate rule for estimating probabilities of dynamical transitions adiabatically. The two examples that we have discussed in this Section should show how useful the Liouvillian perspective on dynamical control can be, even just qualitatively. The intuitive picture of incompressible phase space flowing into or out of growing or shrinking regions is both simple and accurate enough to be a useful guide in designing control protocols, in engineering Hamiltonian systems to achieve control tasks, or in reverse-engineering natural Hamiltonian systems that achieve some control goals, in order to understand how they work.

\section{V. Discussion: Liouville control}
\subsection{Blind control of fast systems}
For the general task of Hamiltonian control, with options to engineer the functional form of coupling terms as well as tune and vary parameters, Liouville control is a uniquely powerful tool. The incompressibility of phase space is built into the very definition of phase space: it is more universal than any particular force law no matter how fundamental, and more general even than energy conservation itself. The concept of Liouville control is to exploit this inherent feature of physical time evolution to ensure that an acceptable fraction of initial conditions \textit{must} evolve into the target region of phase space---no matter what happens.

Liouville's theorem applies to all Hamiltonian systems, but our analysis throughout this paper has considered Hamiltonians to which adiabatic theory (including neo- or post-adiabatic theory) may be applied, because they generate dynamics which is fast compared to the time scale over which the Hamiltonian itself is changing. Applying a Liouvillian perspective to controlling adiabatic systems is not an arbitrary focus, however, because if Liouville control has a killer application it is likely to be in the control of fast systems. The adiabatic approximations which apply in such cases allow computation of relevant phase space volumes just by studying instantaneous Hamiltonians, without actually solving the equations of motion. Under these adiabatic conditions Liouville control can be powerful indeed. As we saw in our previous Section, the design constraints that are implied by the Liouvillian need for increasing separatrix area may be absolutely required for effective control, and yet their necessity may not be apparent at all until without the Liouvillian perspective.

If Liouville control works best for controlling fast systems, moreover, then fast systems may also require techniques like Liouville control. We only even speak of a control task, after all, if control as a task is non-trivial because the system in question does not naturally do as we wish. One of the basic reasons why systems naturally elude our control is that they evolve too quickly for us to perceive and adjust. From this point of view it is a great advantage of Liouville control that it is a form of what could be called \emph{blind control}. It does not depend on any monitoring of the state of the system in order to recognize deviations and correct them. In fact it does not even try at all to adjust any individual trajectory. Liouville control is willing to let any individual trajectory elude control, because growing phase space volume ensures that there will be other trajectories that do behave as desired.

\subsection{Spontaneous change in Hamiltonian evolution}
It may even be appropriate to say that Liouville control arranges to have desired dynamical transitions occur \emph{spontaneously}. `Spontaneous' is not a term that is normally used in deterministic dynamics, yet dynamical transitions of fast systems by crossing expanding separatrices would qualify as `spontaneous' in at least two colloquial senses. 

First of all, these post-adiabatic transitions are impossible to predict without knowing the system's state very precisely. The KNH formula rather easily provides a prediction of probabilities, but the dynamical phase into which any given trajectory will finally settle often depends sensitively on dynamical variables that evolve very rapidly. Numerically solving equations of motion like those in this paper, and trying to guess the final phase from initial conditions selected at random, feels very much like trying to tell whether a birthday candle will light at any given touch of the match, or whether a lawnmower motor will start on any given pull of the cord. In practical terms these transitions are spontaneous in the sense of being unpredictable.

They are also spontaneous, however, in the second colloquial sense that they happen without being \emph{forced} to happen by precisely controlling all involved causal factors. The transitions are unpredictable without precise knowledge of microscopic fast variables, but even without control of microscopic fast variables, the transitions occur---at least with probability sufficient that if the process does fail, it pays to simply keep trying. It might take a few pulls on the cord but the motor will start.

\subsection{Microscopic precursors of thermodynamics?}
If we thus compare dynamical transitions in small Hamiltonian systems to spontaneous processes in macroscopic systems, it is natural to ask about the relationship between phase space area increase as required by Liouville control, and entropy increase as required by the Second Law of Thermodynamics. According to statistical mechanics, entropy increase is also to be interpreted as increase of a certain phase space volume. On the other hand, statistical mechanics considers the relevant volume to be that which is ergodically explored by a system in equilibrium. Some of the arguments for applying KNH-like formulas to realistic ensembles of orbits may have invoked principles somewhat like ergodicity, in assuming that all typical ensembles must have similar capture fractions, but the basic requirement of separatrix growth for Liouville control is based precisely on the fact that an orbit which crosses a separatrix \emph{cannot} explore the whole enclosed phase space volume, because the interior of this region is already incompressibly occupied by other orbits. 

To compare post-adiabatic theory with statistical mechanics in further detail would go far beyond the scope of this paper. We close simply by noting that the qualitative resemblance between spontaneous transitions into growing separatrices, and spontaneous changes that increase total entropy, provides some further support for the hypothesis raised in previous work \cite{Gilz_et_al_2016}, that thermodynamics might not emerge from mechanics in the limit of large system size, but rather represent the persistence into the regime of large systems of dynamical constraints that are already present in the post-adiabatic mechanics of small systems.

\section{Appendix: }
\subsection{Deriving the extended KNH formula}

\subsubsection{Arbitrary `start line' curve as an artificial border}
\begin{figure}[htb]
\centering
\includegraphics[width=.475\textwidth]{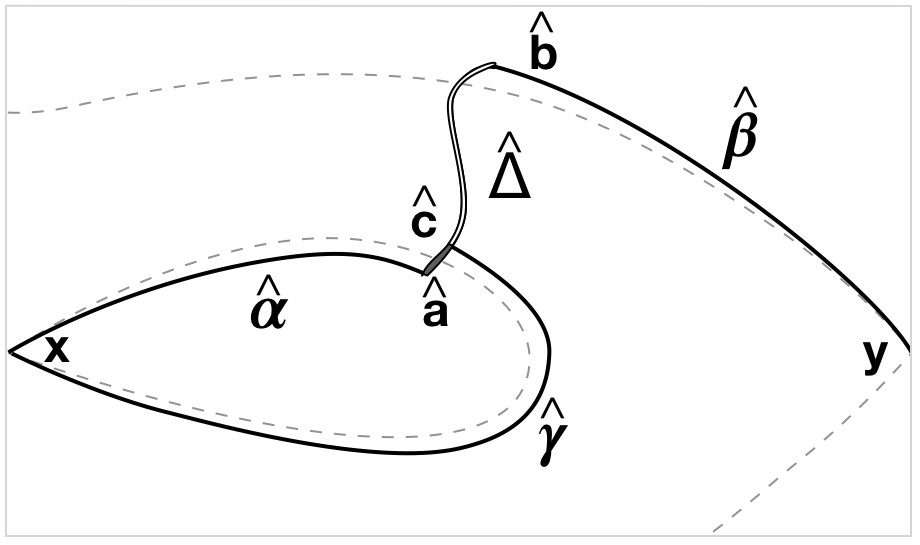}
\caption{\label{FigA1} Curves and points relevant to the extension of the KNH formula. In a region including the instantaneous separatrix (dashed gray contour), an arbitrary curve $\hat{\Delta}$ cuts across system orbits at some time $t$.  The exact orbit $\hat{\alpha}$ is the unique one which reaches $\hat{\Delta}$ at time $t$, having begun at the unstable fixed point $\mathbf{x}$ at some earlier time; the point $\hat{\mathbf{a}}$ is the point at which $\hat{\alpha}$ hits $\hat{\Delta}$. The exact orbits $\hat{\beta}$ and $\hat{\gamma}$ are the unique two which begin on $\hat{\Delta}$ at time $t$, and will eventually approach the unstable fixed points $\mathbf{y}$ and $\mathbf{x}$ respectively. Their starting points on $\hat{\Delta}$ are $\hat{\mathbf{b}}$ and $\hat{\mathbf{c}}$ respectively. All orbits which pass through $\hat{\Delta}$ between $\hat{\mathbf{a}}$ and $\hat{\mathbf{c}}$ will be captured into the closed separatrix, while those which cross between $\hat{\mathbf{c}}$ and $\hat{\mathbf{b}}$ will flow around the closed separatrix into the lower half of phase space.}
\end{figure}

We assume a slowly time-dependent Hamiltonian $H(\mathbf{r},t)$ whose separatrices resemble those of $H_{2}$ \eqref{Htrans} but which is otherwise general; we further assume that we are in a canonical representation such that the unstable fixed points $\mathbf{x}$ and $\mathbf{y}$ do not move over time. Our derivation will refer first of all to Fig.~\ref{FigA1}, which shows the same region of phase space that we showed in our main text's Fig.~\ref{Fig6}, but now with attention on certain curves and points defined by the system's \emph{exact} time evolution, rather than the adiabatic separatrices shown in Fig.~\ref{Fig6}. Since these new points and curves will be approximated by the adiabatic points and curves, they are labelled with the same symbols as their corresponding adiabatic counterparts in Fig.~\ref{Fig6}, but now with circumflex accents.

We start from the fact that the exact orbits follow the instantaneous energy contours approximately, although not exactly. We can therefore consider any arbitrary curve $\hat{\Delta}$ which cuts across a range of orbits that all flow through the neighborhood of an instantaneous separatrix at some arbitrary time $t$. See Fig.~\ref{FigA1}. This curve $\hat{\Delta}$ will let us pose a well-defined capture probability question even though we only have two phase space regions bordered by a separatrix: we will ask what fraction of the orbits which pass through $\hat{\Delta}$ at time $t$ will eventually end up captured inside the inner separatrix (the inner dashed loop in Fig.~\ref{FigA1}). This fraction will turn out to be related in an understandable way to the KNH formula, if $\hat{\Delta}$ is considered as an additional artificial border that splits the infinitely wrapping phase space corridor into an $A_{0}$ and an $A_{-}$, as shown in Fig.~\ref{Fig6} of our main text.

On any such curve $\hat{\Delta}$ we can uniquely identify three important points, denoted $\hat{\mathbf{a}}$, $\hat{\mathbf{b}}$, and $\hat{\mathbf{c}}$ in Fig.~\ref{FigA1}. Point $\hat{\mathbf{a}}$ is the end point on $\hat{\Delta}$ at time $t$ of the trajectory $\hat{\alpha}$ which reaches $\hat{\Delta}$ at time $t$ having started earlier at the unstable fixed point $\mathbf{x}$, which is itself time-independent by the construction of \cite{Tennyson}. This trajectory is unique, and the earlier time at which it began as $\mathbf{x}$ will be denoted $t_{a}<t$. Points $\hat{\mathbf{b}}$ and $\hat{\mathbf{c}}$ are initial points at time $t$ of the unique trajectories $\hat\beta$ and $\hat\gamma$ that will later end at the unstable fixed points $\mathbf{y}$ and $\mathbf{x}$ respectively, reaching them at times $t_{b},t_{c}>t$. If our system is really like $H_{2}$ then $\mathbf{x}$ and $\mathbf{y}$ may be identified as the same point, but we will still need to allow $H(\mathbf{x},t)\not= H(\mathbf{y},t)$ because Hamiltonians like $H_{2}$ are multiply valued. (Every wrapping of $\phi\to\phi-2\pi$ bringing a shift $H_{2}\to H_{2}-2\pi\dot{\alpha}$.)

As will be clear from Fig.~\ref{FigA1}, if $\hat{\mathbf{c}}$ lies above $\hat{\mathbf{a}}$ on $\hat\Delta$, as it does in the Figure, then all points on $\hat\Delta$ between $\hat{\mathbf{a}}$ and $\hat{\mathbf{c}}$ will eventually be trapped inside the closed separatrix, while those between $\hat{\mathbf{c}}$ and $\hat{\mathbf{b}}$ will flow around the closed separatrix into the lower half of phase space. (If $\hat{\mathbf{c}}$ lies below $\hat{\mathbf{a}}$ then no orbits will be captured; the capture probability is exactly zero and we do not need to consider this case any further.) The range of points along $\hat{\Delta}$ between $\hat{\mathbf{a}}$ and $\hat{\mathbf{b}}$ represent all those which will `decide' whether to enter $A_{+}$ or $A_{-}$ around the time $t$:
points further inside the inner separatrix than $\hat{\mathbf{a}}$ have already been trapped in the separatrix for long enough to orbit around inside it, while points beyond $\hat{\mathbf{b}}$ will wrap around in the periodic $\phi$ co-ordinate to approach the separatrix again at some time significantly later than $t$. To estimate the probabilities of orbits through $\hat\Delta$ being captured at time $t$, therefore, we can restrict our attention to the portion of $\hat\Delta$ between $\hat{\mathbf{a}}$ and $\hat{\mathbf{b}}$. For the orbits outside this range, the decision on capture has either already been made or will not yet be made for some time to come.

\subsubsection{Exact capture fraction}

In every Hamiltonian system the flux $\Phi_{\hat{\mathbf{a}},\hat{\mathbf{c}}}\von{t}$ at time $t$ through a curve in phace space that connects point $\hat{\mathbf{a}}$ with $\hat{\mathbf{b}}$ equals $H\von{\hat{\mathbf{c}},t}-H\von{\hat{\mathbf{a}},t}$. The fraction of phase space passing through $\hat\Delta$ between $\hat{\mathbf{a}}$ and the intermediate point $\hat{\mathbf{c}}$ in the interval $dt$ around $t$, and the total flux between $\hat{\mathbf{a}}$ and the other endpoint $\hat{\mathbf{b}}$, is therefore the exact capture probability we seek for the curve $\hat\Delta$ at time $t$:
\begin{equation}\label{Pdelta}
\mathcal P_{+}(t)=\frac{\Phi_{\hat{\mathbf{a}},\hat{\mathbf{c}}}}{\Phi_{\hat{\mathbf{a}},\hat{\mathbf{b}}}}=\frac{H(\hat{\mathbf{c}},t)-H(\hat{\mathbf{a}},t)}{H(\hat{\mathbf{b}},t)-H(\hat{\mathbf{a}},t)}\;.
\end{equation}
While compact, this expression is not a useful substitute for the KNH formula because determining the points $\hat{\mathbf{a}}$, $\hat{\mathbf{b}}$ and $\hat{\mathbf{c}}$ exactly requires solving the equations of motion and the merit of the KNH formula was to make a prediction without having to do that.

\subsubsection{A Hamiltonian identity for flux through a curve}
For any Hamiltonian system with Hamiltonian $H$ and any open curve $S$ parametrized $\mathbf{r}(s)=(q(s),p(s))$ in phase space, the phase space measure of system points evolving through the curve within any short time $dt$ is $dt$ times the flux through $S$ of the phase space flow field
\begin{equation}
\dot{\mathbf{r}}\equiv\left(\frac{\partial H}{\partial p},-\frac{\partial H}{\partial q}\right)\Big\vert_{q,p}
\end{equation}
which represents the system's time evolution. Directly from Hamilton's equations we find that this flux is identically equal to the difference between the values of the Hamiltonian $H$ at the endpoints $\mathbf{x}$ and $\mathbf{y}$ of the curve $S$:
\begin{eqnarray}\label{PhiS}
\Phi_{S}(t)&=&\int_\mathbf{x}^{\mathbf{y}}\!ds\,\Bigl(\partial_{t}q,\partial_{t}p\Bigr)\cdot\Bigl(\partial_{s}p,-\partial_{s}q\Bigr)\nonumber\\
&=&\int_\mathbf{x}^{\mathbf{y}}\!ds\,\left(\frac{\partial H}{\partial p}\frac{\partial p}{\partial s} + \frac{\partial H}{\partial q}\frac{\partial q}{\partial s}\right)\nonumber\\
& \equiv& \int_\mathbf{x}^{\mathbf{y}}\!ds\,\frac{dH}{ds} \equiv H(\mathbf{y},t)-H(\mathbf{x},t)\;.
\end{eqnarray}

\subsubsection{Exact capture fraction in terms of integrals along orbits}
The time evolution of the Hamiltonian itself along an exact orbit obeys $dH/dt = \partial H/\partial t$. Trivially, therefore, we can write
\begin{eqnarray}\label{contours1}
H(\hat{\mathbf{a}},t) &=& H(\mathbf{x},t_{a})+\int_{t_{a}}^{t}\!dt'\,\partial_{t'}H(\mathbf{r}_{\hat\alpha}(t'), t')\\
&\equiv& H(\mathbf{x},t)+\int_{t_{a}}^{t}\!dt'\,\partial_{t'}[H(\mathbf{r}_{\hat\alpha}(t'), t')-H(\mathbf{x},t')]\nonumber
\end{eqnarray}
as well as similar expressions for $H(\hat{\mathbf{b}},t)$ and $H(\hat{\mathbf{c}},t)$, involving integrals along the curves $\hat\beta$ and $\hat\gamma$.
We thus obtain an equivalent expression for the capture probability that is less compact than (\ref{Pdelta}) but will turn out to be more easily computable:
\begin{widetext}
\begin{equation}\label{Pdelta2}
\mathcal P_{+}(t)=\frac{-\int\limits_{\hat\gamma}\!dt'\,\partial_{t'}[H(\mathbf{r}(t'),t')-H(\mathbf{x},t')]-\int\limits_{\hat\alpha}\!dt'\,\partial_{t'}[H(\mathbf{r}(t'),t')-H(\mathbf{x},t')]}{H(\mathbf{x}',t)-H(\mathbf{x},t)-\int\limits_{\hat\beta}\!dt'\,\partial_{t'}[H(\mathbf{r}(t'),t')-H(\mathbf{x}',t')]-\int\limits_{\hat\alpha}\!dt'\,\partial_{t'}[H(\mathbf{r}(t'),t')-H(\mathbf{x},t')]}\;.
\end{equation}
\end{widetext}

\subsubsection{Adiabatic approximation as integrals along energy contours} 
\begin{figure}[htb]
\centering
\includegraphics[width=.475\textwidth]{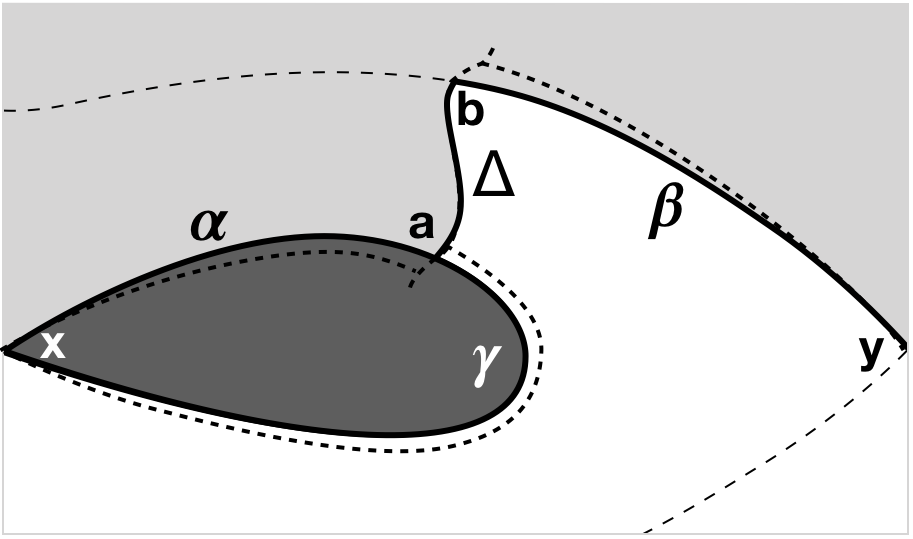}
\caption{\label{FigA2} The adiabatic regions, curves, and points from Fig.~\ref{Fig6} in the main text, with the exact trajectories $\hat{\alpha}$, $\hat{\beta}$ and $\hat{\gamma}$ from Fig.~\ref{FigA1} above shown with dashed lines for comparison. The adiabatic point $\mathbf{a}$ lies on the inner separatrix close to both the exact points $\hat{\mathbf{a}}$ and $\hat{\mathbf{c}}$ (it is in general between them), while the adiabatic point $\mathbf{b}$ is on the outer separatrix and close to the exact point $\hat{\mathbf{b}}$.}
\end{figure}

This expression (\ref{Pdelta2}) for the capture probability has assumed nothing about adiabaticity, but we can now begin approximating it systematically using the fact that the explicit time dependence of $H(\mathbf{r},t)$ is slow. The integrals over $t'$ are of the partial derivative of $H$ with respect to $t'$, and are hence automatically small in the adiabatic limit. By discarding only higher-order post-adiabatic corrections, we can approximate $\partial_{t'}H(\mathbf{r},t')\to\partial_{t} H(\mathbf{r},t)$ in these integrands, since the exact integrands $\partial_{t'}H(\mathbf{r},t')$ will change only slightly over the curves $\hat\alpha$, $\hat\beta$, and $\hat\gamma$. We can also exploit the fact that in the adiabatic limit the exact time evolution trajectories are close to the adiabatic orbits under the instantaneous Hamiltonian $H(\mathbf{r},t)$, by replacing the exact curves $\hat\alpha$, $\hat\beta$, $\hat\gamma$ with their adiabatic approximations ${\alpha}$, ${\beta}$, ${\gamma}$ that are each portions of a separatrix contour. See Fig.~\ref{FigA2}.

Thus approximated to first order in the small adiabaticity parameter, we have
\begin{widetext}
\begin{eqnarray}\label{Pdelta3}
	\mathcal P_+(t)\doteq
\frac{-\int\limits_{{\gamma}}\!ds\,\partial_{t}[H(\mathbf{r}(s),t)-H(\mathbf{x},t)]-\int\limits_{{\alpha}}\!ds\,\partial_{t}[H(\mathbf{r}(s),t)-H(\mathbf{x},t)]}
{H(\mathbf{x}',t)-H(\mathbf{x},t)-\int\limits_{{\beta}}\!ds\,\partial_{t}[H(\mathbf{r}(s),t)-H(\mathbf{x}',t)]-\int\limits_{{\alpha}}\!ds\,\partial_{t}[H(\mathbf{r}(s),t)-H(\mathbf{x},t)]}\;.
\end{eqnarray}
\end{widetext}
where the separatrix contours ${\alpha}$, ${\beta}$ and ${\gamma}$ are parametrized such that
\begin{eqnarray}\label{scond}
	\partial_{s}q &=& \partial_{p}H\nonumber\\
	\partial_{s}p &=& -\partial_{q}H
\end{eqnarray}
so that their $s$-integrals correctly approximate the $t'$-integrals along the exact evolution curves $\hat\alpha$, $\hat\beta$ and $\hat\gamma$.

\subsubsection{Identities for separatrices}
We then note that the instantaneous closed and open separatrices are defined as the contours $H(\mathbf{r},t)=H(\mathbf{x},t)$ and $H(\mathbf{r},t)=H(\mathbf{y},t)$, respectively---so the energy of \textit{any} point on the open separatrix at time $t$ is $H(\mathbf{y},t)$, and $H(\mathbf{x},t)$ is the energy of any points on the inner separatrix. If we therefore consider the evolution flux through the curve ${\Delta}$, which is the portion of our arbitrary curve $\hat\Delta$ between the instantaneous separatrices at $t$, we can apply the Hamiltonian identity (\ref{PhiS}) from above to see that
\begin{eqnarray}\label{}
	H(\mathbf{y},t)-H(\mathbf{x},t) \equiv H({\mathbf{\gamma}},t)-H({\mathbf{\alpha}},t)\equiv \Phi_{{\Delta}}(t)\;.
\end{eqnarray}

From the general definition of a separatrix $\Sigma$ at time $t$ as a contour $\mathbf{r}(s,t)$ such that
\begin{eqnarray}\label{Sigma}
	H(\mathbf{r}(s,t),t)-E(t)=0 
\end{eqnarray}
we can also differentiate with respect to $t$ to obtain
\begin{eqnarray}\label{tdif}
	0&=&\frac{\partial H}{\partial q}\frac{\partial q}{\partial t}+\frac{\partial H}{\partial p}\frac{\partial p}{\partial t}+\partial_{t}[H-E]\nonumber\\
&=&-\frac{\partial p}{\partial s}\frac{\partial q}{\partial t}+\frac{\partial q}{\partial s}\frac{\partial p}{\partial t}+\partial_{t}[H-E]
\end{eqnarray}
as the equation which determines the change in time of the separatrix contour $\mathbf{r}(s,t)$. We have used here the canonical parametrization condition (\ref{scond}) for the separatrix $\mathbf{r}(s,t)$. This implies immediately that the rate of growth of the area $\mathbf{a}$ enclosed by a separatrix $\Sigma$ is
\begin{eqnarray}\label{dotAgen}
	\dot{A}(t) &\equiv& \int\limits_{\Sigma}\!ds\,\mathbf{\hat{n}}(s)\cdot\partial_{t}\mathbf{r}(s,t)\nonumber\\
&\equiv&\int\limits_{\Sigma}\!ds\,\left(\frac{\partial p}{\partial s}\frac{\partial q}{\partial t}-\frac{\partial q}{\partial s}\frac{\partial p}{\partial t}\right)\nonumber\\
&=& \int\limits_{\Sigma}\!ds\,\partial_{t}[H(\mathbf{r}(s,t),t)-E(t)]\;.
\end{eqnarray}

\subsubsection{The extended KNH formula}
Returning to our particular $H$ with its arbitrary curve $\hat\Delta$ and closed and open separatrices, therefore, we can compare (\ref{Pdelta3}) with (\ref{dotAgen}) to conclude that up to first order in the small adiabatic parameter we have \begin{eqnarray}\label{P+0}
	\mathcal P_{+}(t) = \frac{\dot{A}_{+}(t)}{\Phi_{-}(t)-\dot{A}_{0}(t)}
\end{eqnarray}
if we define $A_{0}$ and $A_{-}$ to be the upper and lower halves of the infinitely winding corridor, as divided by ${\Delta}$. The flux $\Phi_{-}\equiv \Phi_{{\Delta}}$ is the instantaneous flux at time $t$ of $\dot{\mathbf{r}}$ through ${\Delta}$, as described in the abbreviated derivation that we gave in our main text.

\subsubsection{Independence of $\Delta$ to leading order}

\begin{figure}[t!h]
\centering
\includegraphics[width=.475\textwidth]{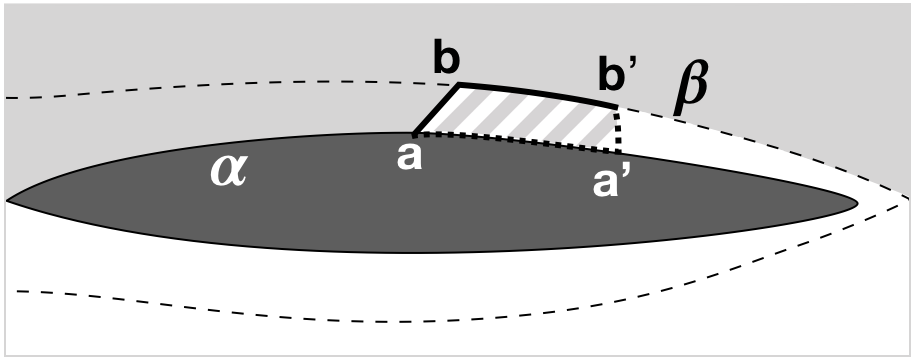}
\caption{\label{FigA3} Example sketch to show the unimportance of the precise location of $\Delta$ in the limit where the two separatrices run close enough to each other  that the flux $\Phi_{-}$ between them is on the order of the adiabatic small parameter. The only difference created by choosing the arbitrary border curve to be $\Delta'$ between $\mathbf{a}'$ and $\mathbf{b}'$, instead of $\Delta$ between $\mathbf{a}$ and $\mathbf{b}$, is whether the $\dot{A}_{0}$ term in (\ref{P+0}) includes an area change integral along $\beta$ between $\mathbf{b}$ and $\mathbf{b}'$, or whether this portion of the total integral is replaced with an integral along $\alpha$ between $\mathbf{a}$ and $\mathbf{a}'$. The integrand is of first order in the adiabatic small parameter anyway; and since the two alternative contours only differ by a displacement of this order as well, the two possible contributions to $\dot{A}_{0}$ differ only at \emph{second} order in the adiabatic small parameter.}
\end{figure}

As we noted in our main text, $\mathcal{P}_{+}$ as given by (\ref{P+0}) appears to depend on exactly where the arbitrary curve $\hat{\Delta}$ has been drawn, but in fact this apparent dependence is illusory. The term in (\ref{P+0}) than depends on $\Delta$ is \emph{not} $\Phi_{-}$, because it follows from (\ref{PhiS}) above that $\Phi_{-}(t)\equiv H(\mathbf{y},t)-H(\mathbf{x},t)$ is the same for \emph{all} ${\Delta}$ which run between the two separatrices. Neither does the $\dot{A}_{+}$ numerator depend on $\Delta$, since it is defined by the closed separatrix. The only dependence on $\Delta$ in $\mathcal P_{+}$ according to (\ref{P+0}) is in $\dot{A}_{0}$, since the choice of where to draw $\Delta$ determines how much of the border of $A_{0}$ is the contour ${\gamma}$, running along the outer separatrix, and how much of it is the contour ${\alpha}$ running along the inner separatrix. But the $-\dot{A}_{0}$ term in the denominator of (\ref{P+0}) is always of first order in the adiabatic small parameter, while in general the flux $\Phi_{-}$ is of zeroth order. So in general the $\Delta$-dependence of $\mathcal P_{+}$ in (\ref{P+0}) is only a higher-order post-adiabatic correction, which must always be added to this leading-order formula, anyway.

We might therefore say that we should drop the $-\dot{A}_{0}$ term from the $\mathcal P_{+}$ denominator, and retain only $\Phi_{-}$; but the special case can still arise, as indeed it does in our $H_{2}$ as derived from $H_{1}$, where $\Phi_{-}$ is also of first order in the adiabatic slowness parameter. In this special case we need to include $-\dot{A}_{0}$ in order to maintain a leading-order result. In this special case, however, the two contours ${\gamma}$ and ${\alpha}$ are necessarily very close to each other, since their energies differ only on the order of the small adiabatic parameter. See Fig.~\ref{FigA3}. The differences in $\dot{A}_{0}$ due to different placements of $\Delta$ will therefore be only of \emph{second} order in the adiabatic small parameter. The extended KNH formula (\ref{P+0}) can therefore be used as written, for any convenient choice of curve $\Delta$, to give a leading-order adiabatic estimate of the capture probability for any size of $\Phi_{-}$.

\end{document}